\documentclass[aps,pra,reprint,10pt,superscriptaddress]{revtex4-1}
\usepackage{amsfonts,bbm,mathtools}
\usepackage{graphicx}
\usepackage[usenames,dvipsnames]{color}
\usepackage[colorlinks,breaklinks,linkcolor=magenta,citecolor=blue,urlcolor=blue]{hyperref}
\usepackage{tikz}

\newcommand{\cC}{\mathcal{C}}
\newcommand{\cD}{\mathcal{D}}
\newcommand{\cE}{\mathcal{E}}
\newcommand{\cF}{\mathcal{F}}
\newcommand{\cG}{\mathcal{G}}
\newcommand{\cH}{\mathcal{H}}
\newcommand{\cI}{\mathcal{I}}

\newcommand{\cR}{\mathcal{R}}
\newcommand{\cS}{\mathcal{S}}

\newcommand{\cU}{\mathcal{U}}
\newcommand{\cV}{\mathcal{V}}
\newcommand{\cW}{\mathcal{W}}
\newcommand{\cX}{\mathcal{X}}
\newcommand{\cY}{\mathcal{Y}}
\newcommand{\cZ}{\mathcal{Z}}

\newcommand{\sG}{\mathsf{G}}
\newcommand{\sId}{\mathsf{Id}}
\newcommand{\sP}{\mathsf{Pauli}}
\newcommand{\sC}{\mathsf{Clif}}
\newcommand{\sR}{\mathsf{RClif}}

\newcommand{\fp}{\mathfrak{p}}
\newcommand{\fh}{\mathfrak{h}}
\newcommand{\fs}{\mathfrak{s}}

\newcommand{\upi}{\mathrm{i}}
\newcommand{\upe}{\mathrm{e}}
\newcommand{\upd}{\mathrm{d}}

\newcommand{\id}{\mathbbm{1}}
\newcommand{\sinc}{\mathrm{sinc}}
\newcommand{\diag}{\mathrm{diag}}
\newcommand{\tw}[1]{\mathcal{T}_{#1}}
\newcommand{\tr}{\mathrm{tr}}
\newcommand{\HS}{\mathrm{H}_\mathrm{S}}

\newcommand{\up}[1]{^{(#1)}}
\newcommand{\avg}[1]{\langle{#1}\rangle}

\newcommand{\zG}{z_{\sG}}
\newcommand{\Gz}{\Gamma^{(0)}}
\newcommand{\Go}{\Gamma^{(1)}}

\newcommand{\trB}{\tr_\mathrm{B}}
\newcommand{\trS}{\tr_\mathrm{S}}
\newcommand{\mb}{\mathrm{b}}
\newcommand{\mI}{\mathrm{I}}

\newcommand{\Dt}{\Delta t}

\begin{document}
\title{Randomized benchmarking in the presence of\\ time-correlated dephasing noise}
\author{Jiaan Qi}
\affiliation{Department of Physics, National University of Singapore}
\author{Hui Khoon Ng}
\affiliation{Yale-NUS College, Singapore}
\affiliation{Centre for Quantum Technologies, National University of Singapore}
\affiliation{MajuLab, CNRS-UCA-SU-NUS-NTU International Joint Research Unit, Singapore}
\date{\today}

\begin{abstract}
Randomized benchmarking has emerged as a popular and easy-to-implement experimental technique for gauging the quality of gate operations in quantum computing devices. A typical randomized benchmarking procedure identifies the exponential decay in the fidelity as the benchmarking sequence of gates increases in length, and the decay rate is used to estimate the fidelity of the gate. That the fidelity decays exponentially, however, relies on the assumption of time-independent or static noise in the gates, with no correlations or significant drift in the noise over the gate sequence, a well-satisfied condition in many situations. Deviations from the standard exponential decay, however, have been observed, usually attributed to some amount of time correlations in the noise, though the precise mechanisms for deviation have yet to be fully explored. In this work, we examine this question of randomized benchmarking for time-correlated noise---specifically for time-correlated dephasing noise for exact solvability---and elucidate the circumstances in which a deviation from exponential decay can be expected.
\end{abstract}

\maketitle

\section{Introduction}

Characterizing the noise afflicting a quantum system is an important part of diagnosing the flaws, and thereby mitigating or removing the negative effects, in a physical component of a quantum computing device. Quantum process tomography \cite{Chuang1997, Poyatos1997,DAriano2001,Altepeter2003} allows the complete reconstruction of the noise process, though one has to account for the effects of preparation and measurement errors unavoidable in the tomography procedure. This full reconstruction rapidly becomes very resource-expensive as the system size grows. One often also does not need the full process to understand the problems with the device. Partial noise characterization methods \cite{Mohseni2006,Emerson2007,Shabani2011,Schmiegelow2011}, sensing only certain features of the noise but with significantly reduced resource demands, have thus become very popular.

A common approach to partial noise characterization is randomized benchmarking (RB) \cite{Knill2008,Magesan2011,Magesan2012,Moussa2012}. The original idea is from Emerson et al.~\cite{Emerson2007}, where a group symmetrization, or group twirling, procedure is used to to ``coarse-grain" the quantum process to isolate a few useful group-averaged parameters of the full process description. In its simplest form, the RB protocol makes use of the Clifford group, and assumes that the underlying noise process is a static, gate-independent map. The formal twirling operation over the Clifford group can be approximated by an easily implemented---hence its popularity---procedure of applying random sequences of Clifford group elements and then averaging over a sufficiently large number of sequences. The group twirling results in a depolarizing channel, in place of the original noise process, with a strength determined by, and characteristic of, the original noise. This depolarizing strength can be estimated from the exponential (as predicted by standard RB analysis) decay rate of the fidelity of the output state to the input, for increasing sequence lengths. 

Apart from the Clifford group, RB-like procedures using other groups, like the Pauli group, the dihedral group, and the real Clifford group have also been explored \cite{Emerson2007,Carignan2015,Brown2018,Hashagen2018}, yielding other group-averaged noise features. Going beyond standard RB, gate-dependent noise can be accommodated to some extent. For example, one can use the interleaved version of RB \cite{Magesan2012interleaved} which focuses on the noise in a particular gate of interest. Weak gate dependence can also be incorporated with a modified analysis and interpretation of the RB measurements \cite{Wallman2017,Proctor2017,Qi2019,Carignan2018}.

RB in the presence of time-dependent noise, however, is less well understood. In this case, the noise associated with the same gate can depend on when it is applied, and also on the sequence of gates that have occurred before it. Time correlations in the noise are certainly expected in a real quantum device, though the correlations may be weak. Most RB experiments do observe the exponential fidelity decay curves as predicted by standard RB for static noise; however, there are experiments that report observations of non-exponential decay \cite{Ryan2009,Veldhorst2014,Park2016,Kawakami2016}, suggesting a deviation from the usual assumption of static, time-independent noise. Some have attributed this to time correlations in the noise. Ref~\cite{Fogarty2015} used a model where the noise fidelity value is fixed during each RB sequence, but varies slowly across different sequences. With two distinct fidelity decay rates, they were able to recover a non-exponential RB decay, relevant for systems with low-frequency noise. Refs.~\cite{Ball2016, Mavadia2018}, on the other hand, examined time-correlated dephasing noise by modeling it with random unitary rotations on $z$-axis with a specified autocorrelation profile, and converted the averaging in the RB procedure to a random-walk problem. They found that time correlations in the noise cause the RB sequence fidelity distribution function to differ significantly from when correlations are absent.
Numerical simulations of the RB procedure, however, suggest that one should still observe the decay curve as predicted by standard RB analysis even in the presence of time-correlations in the noise \cite{Epstein2014}.

These past observations and studies invite the question of what features in the time-correlated noise would cause deviation of the RB procedure from the expected exponential decay behavior. If time-correlated noise can indeed lead to a slower RB fidelity decay, we must understand the circumstances under which such a situation can arise, as a caution against over-estimation of the quality of the gate operations being characterized. 

In this work, we seek a better understanding of the behavior of RB in the presence of time-correlated noise. We consider, within a single unified treatment, the standard version with the Clifford group, as well as other group-twirling options. Here, we examine an exactly solvable model, that of dephasing noise. While this restriction to pure dephasing noise limits the applicability of our conclusions to general experimental settings, an exact solution allows us to focus our attention on deviations arising solely from the time correlations in the model, rather than due to approximations in the solution. 

Below, we first begin in Sec.~\ref{sec:Prelim} with some preliminary discussion that sets the stage and explains the basic concepts necessary for understanding the remainder of the paper. Sections \ref{sec:class} and \ref{sec:quant} examine the classical and quantum dephasing noise models, respectively, deriving the behavior of RB for the two kinds of noise models. In Sec.~\ref{sec:discussion}, we elucidate the necessary assumptions for which the RB description found in Secs.~\ref{sec:class} and \ref{sec:quant} holds, and in the process discover the circumstances leading to non-exponential decay of the RB fidelity.

\section{Preliminaries}\label{sec:Prelim}
Let us first lay down the basic concepts needed for our discussion below. We will restrict our quantum information processing system to a single qubit, the most common situation for RB experiments.

\subsection{Noisy gates}\label{subsec:noisyGates}
Ideal gate operations are unitary maps, i.e., a gate $\cG$ acts on the state (density operator) of a quantum system as $\cG(\,\cdot\,)=G(\,\cdot\,)G^\dagger$ for some unitary operator $G$. The noisy implementation $\widetilde \cG$ of the ideal gate $\cG$ is modelled as follows: A noisy gate $\widetilde \cG$ that takes time $\tau$ to complete is modelled by evolution according to dephasing noise by time $\tau$, followed by an instantaneous application of the ideal gate $\cG$ at the end of time $\tau$. 
That the noise is dephasing refers to the fact that the noise acts according to a Hamiltonian with a Pauli $Z$ operator on the system, commuting with the bare system-only Hamiltonian assumed to be $\HS=\frac{1}{2}\omega_0Z$ ($\hbar$ has been set to 1) with $\omega_0$ as the energy splitting of the qubit. Below, we consider classical dephasing noise described by interaction with a classical fluctuating field, as well as quantum dephasing noise where the qubit interacts with a bosonic environment. In both cases, the Hamilton operator describing the physical situation can be written as
\begin{equation}
\mathrm{H}=\mathrm{H}_0+\mathrm{H}_\mathrm{noise}, \quad\textrm{with }\mathrm{H}_\mathrm{noise}=B Z,
\end{equation}
where in $\mathrm{H}_\mathrm{noise}$, the source of the noise on the system, $B$---defined concretely in subsequent sections---represents the classical fluctuating field or the bosonic degrees of freedom accompanying the $Z$ operator on the system. $\mathrm{H}_0$ collects the non-interacting part of the Hamilton operator, containing $\HS$ and possibly $\mathrm{H}_\mathrm{B}$ for the bosonic pure-bath term. 
With this, the noisy gate $\widetilde\cG$, starting at time $t$ and taking time $\tau$, can be written as $\widetilde\cG=\cG\cU(t+\tau,t)$, with $\cU(t',t)$ as the evolution map according to $\mathrm{H}$ from time $t$ to time $t'$.

\subsection{Groups of interest}

We focus our discussion here on three groups of unitary maps used in the literature for group-twirling-based noise characterization: the Pauli group $\sP$, the real Clifford group $\sR$, and the (complex) Clifford group $\sC$ \cite{Calderbank1998}. Only the single-qubit groups concern us here. We first define three groups of qubit operators,
\begin{align}
\fp&\equiv \mathrm{cl}\{\sigma_0\equiv \id, \sigma_1\equiv X, \sigma_2\equiv Y, \sigma_3\equiv Z\},\nonumber\\
\fh&\equiv \mathrm{cl}\{\id,H\},\\ 
\textrm{and}\quad\fs&\equiv \mathrm{cl}\{\id, S, S^2\}\nonumber,
\end{align}
where $\mathrm{cl}$ denotes closure under group (operator) multiplication. Here, $X, Y$, and $Z$ are the usual Pauli operators, $X\equiv |0\rangle\langle 1|+|1\rangle\langle 0|$, $Y\equiv -\upi|0\rangle\langle 1|+\upi|1\rangle\langle 0|$, and $Z\equiv |0\rangle\langle 0|-|1\rangle\langle1|$ (so $|0\rangle$ and $|1\rangle$ are the eigenvectors of $Z$), $H\equiv |+\rangle\langle 0|+|-\rangle\langle 1|$ is the Hadamard operator with $|\pm\rangle\equiv \frac{1}{\sqrt 2}(|0\rangle\pm|1\rangle)$, and $S\equiv \omega|+\rangle \langle 0|-\bar \omega|-\rangle\langle 1|$, with $\omega\equiv \upe^{-\upi\pi/4}$ and $\bar\omega$ is its complex conjugate. Note that $S^\dagger =-S^2$.

With these operator groups, it is straightforward to define the groups of unitary maps of interest. The Pauli group $\sP$ is the group of unitary maps such that $\cG\in\sP$ iff $\cG(\cdot)=G(\cdot)G^\dagger$ for $G\in \fp$. $\sP$ thus comprises 4 distinct elements: $\sP=\{\cI,\cX,\cY,\cZ\}$, where $\cI(\cdot)=\id(\cdot)\id$ is the identity map, $\cX(\cdot)=X(\cdot)X$, etc. We can also represent the four elements of $\sP$ by (diagonal) matrices written using the Pauli operator basis $\{\sigma_\alpha\}_{\alpha=0}^3$, 
\begin{align}
\cI\,&\widehat{=}\,\diag\{1,1,1,1\}, &\cX\,&\widehat{=}\,\diag\{1,1,-1,-1\}, \nonumber\\
\cY\,&\widehat{=}\,\diag\{1,-1,1,-1\},&\cZ\,&\widehat{=}\,\diag\{1,-1,-1,1\}.
\end{align}
Here, the entries of the matrix $M$ representing the map $\cG$ is given by $M_{\alpha\beta}\equiv \tr\{\sigma_\alpha \cG(\sigma_\beta)\}$. To simplify notation, we denote both the map $\cG$ and its matrix representation $M$ in the Pauli operator basis---also known as the transfer matrix of $\cG$---by the same symbol $\cG$.

In a similar manner, the real Clifford group $\sR$ is recognized as the group of unitary maps such that $\cG\in\sR$ iff $\cG(\cdot)=G(\cdot)G^\dagger$ for $G\in \mathrm{cl}\{\fp\cup\fh\}=\mathrm{cl}\{\id,X, Y, Z, H,HX,HY,HZ\}$; $\sR$ has 8 distinct elements. The Clifford group $\sC$ is the group of unitary maps such that $\cG\in\sC$ iff $\cG(\cdot)=G(\cdot)G^\dagger$ for $G\in \mathrm{cl}\{\fp\cup\fh\cup\fs\}$; $\sC$ has 24 distinct elements. Defined this way, it is clear that $\sP\subset\sR\subset\sC$. Conventionally, $\sC$ is defined as the set of all unitary maps that preserves the Pauli operator group $\fp$.
For formal definitions and their properties, we refer the readers to Ref.~\cite{Nebe2001}. 
Elements $\cC\in\sC$ and $\cR\in\sR$ have transfer matrices of the form
\begin{equation}
  \cC = \left(
    \begin{array}{cccc}
    1& 0&0 &0\\ 
    0 &* &* &*\\
    0 &* &* &*\\
    0 &* &* &*
    \end{array}\right) \quad\textrm{and }\quad 
    \mathcal{R} = \left(
    \begin{array}{cccc}
    1& 0& ~0&0\\ 
    0&* & ~0&*\\
    0 & 0&\!\!\pm1&0\\
    0 & *&~0 &*
    \end{array}\right),
\end{equation}
where $*$ indicate nonzero entries that differ for different group elements.

\subsection{Group twirling}
As mentioned in the introduction, group-twirling methods allow us to isolate certain features of interest from the original noise process. The group-twirled process is of a simpler form than the original noise, due to the group averaging. The twirl of a map $\cE$, according to a group $\sG$ of unitary maps, is the following group average,
\begin{equation}\label{eq:twirl}
\tw{\sG}(\cE) \equiv \frac{1}{|\sG|}\sum_{\cG\in\sG} \cG \cE \cG^\dagger.
\end{equation}
Here, $|\sG|$ denotes the number of elements in $\sG$, and $\cG^\dagger$ is the adjoint map of $\cG$, defined for a unitary map $\cG(\,\cdot\,)=G(\,\cdot\,)G^\dagger$ with unitary operator $G$, as
$\cG^\dagger (\,\cdot\,)\equiv G^\dagger (\,\cdot\,)G$.

For our current purpose, we need only to understand how a given qubit map $\cE$ transforms under twirling for our three groups of interest. The calculations are given in Appendix \ref{app:twirling}. Here, we only state the results, written in the transfer-matrix representation: For a single-qubit map $\cE$ with transfer matrix 
\begin{equation}\label{eq:transMat}
\cE = {\left(\begin{array}{cccc}
    a& c_1& c_2&c_3\\ 
    b_1&x_{11} & x_{12}&x_{13}\\
    b_2 &x_{21}&x_{22}&x_{23}\\
    b_3 &x_{31}&x_{32}&x_{33}
    \end{array}\right)},
\end{equation}
$\sP$-twirling eliminates all but the diagonal entries, 
\begin{equation}\label{eq:twirlP}
\tw\sP(\cE)=\textrm{diag}\{a,x_{11},x_{22},x_{33}\};
\end{equation}
twirling with the $\sR$ and $\sC$ groups also eliminate the off-diagonal entries, and further mix the on-diagonal ones,
\begin{align}
\label{eq:twirlR}\tw{\sR}(\cE)\!&= \textrm{diag}\{a,v,x_{22},v\}, ~v\equiv\!\tfrac{1}{2}(x_{11}\!+\!x_{33}),\\
\label{eq:twirlC}\tw{\sC}(\cE)\!&= \textrm{diag}\{a,w,w,w\}, ~w\equiv\!\tfrac{1}{3}(x_{11}\!+\!x_{22}\!+\!x_{33}).
\end{align}

\subsection{Randomized benchmarking}
The definition of group twirling in Eq.~\eqref{eq:twirl} does not lend itself to easy direct implementation. 
Instead, the RB scheme \cite{Knill2008, Magesan2011,Magesan2012,Moussa2012} offers a practical approach to estimating the result of the group twirling. In brief, the RB procedure, for twirling over the group $\sG$, involves the following steps:
\begin{enumerate}
    \item Prepare a pure input state $\psi_0$.
    \item Uniform-randomly choose $m$ elements from $\sG$: $\cG_0,\cG_1,\ldots,\cG_{m-1}$; let $\cG_m\equiv \cG_0^\dagger\cG_1^\dagger \ldots\cG_{m-1}^\dagger$.
    \item Apply the gates $\cG_0,\cG_1,\ldots, \cG_m$, in sequence, on $\psi_0$. This amounts to applying the (noisy) sequence $\cS_{m}=\widetilde\cG_m\widetilde\cG_{m-1}\ldots\widetilde\cG_1\widetilde\cG_0$ on the input state.
    \item Measure the (square of the) fidelity of the output state with the input state: $\tr{\left\{\psi_0 \, \cS_{m}(\psi_0)\right\}}$.
    \item Repeat steps 1-4, with different random choices of $\cG_0,\cG_1,\ldots,\cG_{m-1}$ from $\sG$ for the same $\psi_0$, and compute the average---over the many repetitions---fidelity value. Denote this average value by $F_m$.
    \item Repeat steps 1-5 with different values of $m$, and examine how $F_m$ depends on $m$.
\end{enumerate}

If all the gates are ideal, the sequence $\cS_m$ is simply the identity map, by the definition of $\cG_m$, so $F_m=1$ for all $m$, assuming that there are no state preparation and measurement (SPAM) errors. With SPAM errors, but still ideal gates, $F_m$ will be an $m$-independent constant. With noise in the gates as well, one expects $F_m$ to decrease as $m$ increases, as the errors due to the noisy gates accumulate as the sequences get longer. The deviation of $F_m$ from 1 due to SPAM errors, however, do not scale as $m$, making RB robust against SPAM errors. The dependence of $F_m$ on $m$, as we will see in detail below, is related to specific noise parameters of the gate operations.

To see the link between RB and group twirling, we first note that $F_m= \tr{\left\{\psi_0 \, \avg{\cS_{m}}_{\text{seq}}(\psi_0)\right\}}$, where $\avg{\,\cdot\,}_{\textrm{seq}}$ denotes the averaging over the sequences of length $m$. The properties of this average sequence, itself a quantum map, $\avg{\cS_{m}}_{\text{seq}}$, determine the experimentally accessible $F_m$. We write the noisy gate as $\widetilde\cG_k=\cG_k\cE_k$, for some noise $\cE_k\equiv \cG_k^\dagger \widetilde\cG_k$. The noise $\cE_k$ could in general be both time- and gate-dependent. 
Note that, for our chosen description of noisy gates (Sec.~\ref{subsec:noisyGates}), $\cE_k$ is related to (the relationship is made concrete in our specific examples below) $\cU(t_{k+1},t_k)$ for $\cG_k$ applied at time $t_k$ and taking time $\tau=t_{k+1}-t_k$. Furthermore, letting $\cG_m'\equiv \cG_m$ and defining, consecutively for $k=m-1,m-2,\ldots,0$, $\cG_k'$ by the equation $\cG_k=\cG_{k+1}'^\dagger \cG_k'$, we can write $\avg{\cS_{m }}_{\text{seq}}$ as
\begin{align}
\avg{\cS_{m }}_{\text{seq}} &= \avg{\,\cG_{m} \cE_m  \cdots \cG_{0} \cE_0\,}_{\text{seq}}\\
&=\avg{\cG_m'\cE_m\cG_m'^\dagger\cG_{m-1}'\cE_{m-1}{\cG_{m-1}'}^\dagger\ldots\cG_1'^\dagger \cG_0'\cE_0\cG_0'^\dagger}_\textrm{seq}.\nonumber
\end{align}
Due to the definition of $\cG_m$ in terms of the other $\cG_k$s, we find that $\cG_0'=\cI$. The $\cG_k'$s, for $k=1,\ldots, m$, can then be regarded as an independent set of gates randomly chosen from $\sG$, and the sequence average amounts to independent group averaging over $\sG$ for each $k$. We thus have
\begin{equation}\label{eq:seq}
\avg{\cS_{m }}_{\text{seq}} =    \big\langle \tw{\sG}(\cE_m)\tw{\sG}(\cE_{m-1})\ldots\tw{\sG}(\cE_1)\cE_0\big\rangle_\mathrm{b},
\end{equation}
where the sequence average has been split into separate group twirling operations for each $\cE_k$. An additional $\langle\,\cdot\,\rangle_\mathrm{b}$ symbol has been added above, to remind us that the $\cE_k$s depend generally on the particular sequence it belongs to, so the averaging has to be done correctly; we will revisit this point in detail when we discuss the concrete cases of time-correlated dephasing noise below. Different ``aspects" of $\avg{\cS_m}_\textrm{seq}$ can be measured by choosing different input states $\psi_0$, in turn giving us information about the group-averaged parameters of the gate noise in accordance with Eq.~\eqref{eq:seq}. 

If we assume static, gate-independent noise, as in standard RB analyses, we have $\cE_k\simeq \cE$, a fixed map, so that $\avg{\cS_m}_\textrm{seq}\simeq [\tw{\sG}(\cE)]^m\cE$. Then, the fidelity quantity $F_m=\tr\{\psi_0 [\tw{\sG}(\cE)]^m\cE(\psi_0)\}$ follows an exponential decay law, $Ap^m+B$, for some $m$-independent constants $A$ and $B$. $p$ is the decay parameter that can be related to the depolarizing strength of the twirled $\cE$, and hence characteristic of the noise $\cE$ (see the RB papers for more information on this). By observing how $F_m$ depends on $m$ from the RB data, one can estimate the value of $p$ and hence obtain information about the gate noise. For time-correlated noise, this exponential decay need not follow from Eq.~\eqref{eq:seq}; we are interested in this deviation from the exponential law in our dephasing noise discussion below. 

Note that the noise $\cE_0$ for the first gate can be absorbed as part of the preparation noise, and has no impact on the $m$ dependence of $F_m$. In the following, we thus set $\cE_0=\id$ in Eq.~\eqref{eq:seq}, i.e., we regard the first gate as having no noise, with the noise already accounted for as part of the preparation noise.

\section{Classical dephasing noise}\label{sec:class}
As our first example, we examine the outcome of group twirling for a single-qubit system under the influence of a fluctuating classical field. This situation is described by the Hamilton operator,
\begin{equation}
  \mathrm{H}=\tfrac{1}{2}\omega_0 Z + \tfrac{1}{2} b(t) Z,
\end{equation}
where $\omega_0$ is the energy splitting of the qubit. The time-dependent variable $b(t)$ is a real-valued random number, reflecting the instability in the energy splitting induced by unwanted environmental perturbations. We assume that $b(t)$ is a zero-mean, i.e., $\avg{b(t)}_b=0$, and stationary Gaussian process. Here, $\avg{\,\cdot\,}_b$ denotes the average over the random process. Zero-mean stationary Gaussian processes are very well-studied objects \cite{Brockwell2016} with many useful properties. Here we need only three specific properties: (1) the $n$-point correlation function $\langle b(t_1) b(t_2)\ldots b(t_n)\rangle_b$, with odd $n$, vanishes; (2) the two-point correlation function
$\langle b(t) b(t')\rangle_b\equiv C(t-t')$ depends only on the time difference (stationarity); (3) for $\phi(t,t')\equiv \int_{t'}^t \upd\tau\, b(\tau)$,
\begin{equation}\label{eq:gauss-avg}
\left\langle \upe^{i\phi} \right\rangle_b = \upe^{-\frac{1}{2}\avg{\phi^2}_b}.
\end{equation}

In the following, we work in the interaction picture defined by $\mathrm{H}_0=\frac{1}{2}\omega_0 Z$. This simply amounts to us computing the fidelity $F_m$ with the interaction-picture version of $\psi_0$.
We consider the dynamical map that takes the qubit state from time $t'$ to time $t$.
In the transfer-matrix representation, the (interaction-picture) unitary evolution for some given process $b(t)$ is
\begin{equation}\label{eq:UClass}
\cU_\mI(t,t')=\left(
\begin{array}{cccc}
1&0&0&0\\0&\cos\phi(t,t')&\sin\phi(t,t')&0\\
0&-\sin\phi(t,t')&\cos\phi(t,t')&0\\0&0&0&1
\end{array}\right),
\end{equation}
where the phase $\phi(t,t')$ as defined above. In the absence of the RB pulses, the evolution of the qubit from time $t'$ to $t$, averaged over $b(t)$, is described by a dephasing channel, namely, the transfer matrix $\cE_{\mathrm{free}}(t,t')=\langle \cU_\mI(t,t')\rangle_b=\mathrm{diag}\{1,\upe^{-\Gamma},\upe^{-\Gamma},1\}$, with $\Gamma \equiv \frac{1}{2}\avg{\phi^2}_b$

With RB, following Eq.~\eqref{eq:seq} and \eqref{eq:UClass}, we have, after some straightforward algebra for the three twirling groups of interest,
\begin{equation}\label{eq:twClass}
\langle \cS_{m}\rangle_\mathrm{seq} ={\left\{\begin{array}{ll}
\mathrm{diag}\{1,f_m,f_m,1\},&\sG=\sP\\
\mathrm{diag}\{1,g_m,f_m,g_m\},&\sG=\sR\\
\mathrm{diag}\{1,h_m,h_m,h_m\},&\sG=\sC
\end{array}\right.},
\end{equation}
where three ``decay parameters" are introduced by
\begin{align}
f_m&\equiv {\left\langle\prod_{k=1}^m\cos\phi_k\right\rangle}_b,\nonumber\\
g_m&\equiv {\left\langle\prod_{k=1}^m\tfrac{1}{2}(1+\cos\phi_k)\right\rangle}_b,\nonumber\\
\label{eq:decayPara}\textrm{and}\quad h_m&\equiv {\left\langle\prod_{k=1}^m\tfrac{1}{3}(1+2\cos\phi_k)\right\rangle}_b,
\end{align}
with $\phi_k\equiv \phi(t_{k},t_{k-1})=\int_{t_{k-1}}^{t_{k}}\!\upd\tau\,b(\tau)$. 
Recalling that $\cos\phi_k=\frac{1}{2}(\upe^{\upi\phi_k}+\upe^{-\upi\phi_k})$, we can adopt a unified notation for the decay parameters,
\begin{equation}\label{eq:pmClass}
p_m(\sG) = \frac{1}{|A_{\sG}|^m} \!\!\sum_{\vec a_m \in A_{\sG}^m} 
\!\!\!{\left\langle\upe^{\upi \vec a_m \cdot \vec\phi_m }\right\rangle}_{\!b} \!\!
\equiv {\left\langle\!\left\langle
\upe^{\upi \vec a_m \cdot \vec \phi_m}
\right\rangle_{\!b}\right\rangle
}_{\!\vec a_m},
\end{equation}
where $\vec\phi_m\equiv(\phi_1,\phi_2,\ldots,\phi_m)$ is the collection of phases, and
$\vec a_m\equiv(a_1,a_2,\ldots,a_m)$ is an $m$-tuple. Each $a_k$ runs over the elements in the multiset $A_{\sG}$ of $0$ and $\pm 1$; we define $A_\sP\equiv \{-1,1\}$, $A_\sR\equiv \{0,0,-1,1\}$, and $A_\sC\equiv\{0,-1,1\}$. $\langle\,\cdot\,\rangle_{\vec a_m}$ denotes an average over the $m$-tuples $\vec a_m$. Then, $f_m=p_m(\sP)$, $g_m= p_m(\sR)$  and $h_m= p_m(\sC)$, where the associations between the decay parameters and the groups serve only to remind us of Eq.~\eqref{eq:decayPara}; of course, for $\sG=\sR$, both $f_m$ and $g_m$ appear in Eq.~\eqref{eq:decayPara}.  For convenience later, let us denote the fraction of $0$ elements in $A_\sG$ by $z_\sG$. Note that each $A_\sG$ only has two nonzero elements, so $1-z_\sG=2/|A_\sG|$. These definitions are summarized in Table~\ref{tab:GClass}.
\begin{table}[ht]
\centering
\renewcommand*{\arraystretch}{1.5}
\begin{tabular}{c|r|c|c|c} 
$\sG$\  & $A_\sG$\qquad \ & \ $|A_\sG|$   &\ $\zG$  &decay parameter $p_m(\sG)$  \\[0.5ex] 
\hline\hline
$\sP$ & $\{-1,1\}$ & 2 & $0$ &$f_m$\\ 
\hline
$\sR$ & $\{0,0,-1,1\}$ & 4  & $1/2$&$g_m$ \\
\hline
$\sC$ & $\{0,-1,1\}$ & 3 & $1/3$&$h_m$ \\
\end{tabular}
\caption{The multisets $A_\sG$ for twirling groups $\sP$, $\sR$ and $\sC$ in the classical noise example.}
\label{tab:GClass}
\end{table}

Eq.~\eqref{eq:pmClass} involves two averages: the average over the random process $b$, and the tuple average over $\vec a_m$. The $b$ average can be evaluated with the aid of Eq.~\eqref{eq:gauss-avg}, so that $\langle\upe^{\upi \vec a_m \cdot \vec\phi_m }\rangle_{b}=\exp[-\frac{1}{2}{\langle (\vec a_m\cdot\vec \phi_m)^2\rangle}_b]$, and
\begin{align}
&\frac{1}{2}\bigl\langle (\vec a_m \cdot\vec \phi_m)^2 \bigr\rangle_b
=\!\frac{1}{2}\!\sum_{j,k=1}^m \!\! a_j a_{k}\!\! 
\int_{t_{j-1}}^{t_{j}}\!\!\!\upd\tau\! 
\int_{t_{k\!{-}1}}^{t_{k}}\!\!\!\upd\tau'\,C(\tau-\tau')\nonumber\\
&=\!\frac{1}{2}\!\sum_{j,k=1}^m\!\! a_j a_{k} \!\! 
\int_{\!{-}\infty}^\infty\! \frac{\upd \omega}{2 \pi}\, \widetilde C(\omega) 
{\left[\frac{\sin\!\left(\omega\Delta t/2\right)}{\omega/2}\right]}^2
\upe^{-\upi(j-k)\omega\Delta t}.
\end{align}
In the second line above, we have made the usual assumption that the RB gates are applied at regular time intervals of $\Dt$, so that $t_k=k\Delta t$ for $k=0,1,\ldots, m$. $\widetilde C(\omega)\equiv \int_{-\infty}^{+\infty}\upd t\,\upe^{\upi\omega t}C(t)$, the bath spectral power function, is the Fourier transform of $C(t)$.
Switching to dimensionless variable $x\equiv \omega\Delta t$, defining the dimensionless spectral function as $S(x)\equiv \frac{\Delta t}{2\pi}\,\widetilde C{\left(\frac{x}{\Delta t}\right)}$, and symmetrizing the expression in $j$ and $k$, we can rewrite the above as
\begin{align}\label{eq:sum}
\frac{1}{2}\bigl\langle\bigl(\vec a\cdot\vec \phi\bigr)^2\bigr\rangle_b=\sum_{j,k=1}^ma_ja_{k} \Gamma^{(|j-k|)},
\end{align}
where
\begin{equation}\label{eq:GammaClass}
\Gamma^{(n)}\equiv\frac{1}{2} \int_{-\infty}^\infty\!\! \upd x\, S(x){\left[\sinc{\left(\frac{x}{2}\right)}\right]}^2\cos(nx),
\end{equation}
for nonnegative integer $n$, and $\sinc \,x\equiv \sin x/x$. $\Gamma^{(n)}$, which depends on the nature of the noise only, can be regarded as an ``$n$-step'' decoherence function, with $n$ giving the lapse in time, measured in units of $\Delta t$, in the bath correlation function. The $n=0$ case gives the usual dephasing function of $\cE_{\mathrm{free}}$ found earlier, for time interval $\Dt$.
Finally, then, we can write the general decay parameter as
\begin{align}\label{eq:pmClassFull}
p_m(\sG)=\left\langle\exp{\Biggl[-\!\!\sum_{j,k=1}^m \!a_ja_{k} \,\Gamma^{(|j-k|)}\Biggr]}\right\rangle_{\vec a_m}.
\end{align}

The remaining tuple average is, unfortunately, not easily amenable to exact evaluation. Instead, we argue for truncating the sum in Eq.~\eqref{eq:pmClassFull}, keeping only small $n$ values, before evaluating the tuple average. We assume that the noise on the system is such that 
\begin{equation}\label{eq:hiClass}
1\gg |\Gamma^{(0)}| > |\Gamma^{(1)}|\gg |\Gamma^{(2)}|\gg\cdots.
\end{equation}
Such a hierarchy of the $n$-step decoherence functions can be expected as long as $S(x)$ is slowly varying compared to $1/\Dt$; if so, the multiplication of $S(x)$ in $\Gamma^{(n)}$ by the oscillatory $\cos(nx)$ leads to smaller and smaller integral values as $n$ grows. We will further explore the subtleties and implications of this condition in Sec.~\ref{sec:discussion}. For now, we assume it holds, which permits the truncation of the sum over $k$ and $k'$ in $p_m(\sG)$ [Eq.~\eqref{eq:pmClassFull}], keeping only the terms with small $n$ values \footnote{By examining the expression for $p_m(\sG)$ in Eq.~\eqref{eq:pmClassFull}, summing up all exponential factors over the entire configuration space, we can draw a similarity between $p_m(\sG)$ and the partition function of a finite-length Ising chain with long-range coupling $\Gamma\up{n}$. This similarity was also observed in Ref.~\cite{fong2017}. Here, however, the similarity is purely mathematical, not physical. But the Ising model analogy indeed invites us to truncate the sum in Eq.~\eqref{eq:pmClassFull} to include only ``on-site'' terms $k'=k$ and ``nearest-neighbor interaction'' terms $k-k'=\pm1$.}.  

We begin with the zeroth-order approximation, namely where only the ``time-local" terms with $j=k$ are kept in the sum Eq.~\eqref{eq:pmClassFull}:
\begin{align}
p_m\up{0}(\sG) 
& \equiv {\left\langle\upe^{-\sum_{k=1}^m a_k^2 \Gz }\right\rangle}_{\vec a_m}
= \prod_{k=0}^m {\left\langle \upe^{-a_k^2 \Gz}\right\rangle}_{a\in A_\sG}
\nonumber\\
&=\bigl[\zG + (1-\zG)\upe^{-\Gz} \bigr]^m = [p_1(\sG)]^m.
\end{align}
Note that $p_1(\sG)=p_1^{(0)}(\sG)$, an exact equality rather than an approximation, as there are only time-local terms in the $m=1$ case. For longer sequences,
$p_m^{(0)}(\sG)= [p_1(\sG)]^m$ gives the usual exponential decay behavior observed in standard RB. In the usual RB analysis, this exponential decay arises directly from the assumption that the noise is described by a CPTP map in each time step, with no temporal correlations between time steps. This is consistent with our intuition here of a time-local approximation.

To observe the effect of time correlations, we go beyond the zeroth-order approximation. Keeping now up to the $n=1$ terms in Eq.~\eqref{eq:pmClassFull}, we have the first-order approximation,
\begin{equation}\label{eq:classOrder1}
p_m^{(1)}(\sG)\equiv{\left\langle\upe^{-\sum_{k=1}^ma_k{\left[a_k\Gz+(a_{k+1}+a_{k-1})\Go\right]}}\right\rangle}_{\vec a_m},
\end{equation}
where it is understood that $a_{m+1}$ and $a_{-1}$ are identically zero.
The full calculation of the tuple average, involving the solution of a set of recurrence relations, is described in Appendix \ref{app:class}. Here, we give only the final answer,
\begin{align}\label{eq:pm1Class}
 p_m^{(1)}(\sG)= \frac{p_1\up{1}-\lambda_-}{\lambda_+-\lambda_-}\, \lambda_+^m
 +\frac{\lambda_+-p_1\up{1}}{\lambda_+-\lambda_-}\, \lambda_-^m,
\end{align}
where the two decay parameters are
\begin{equation}
\lambda_\pm  = \frac{1}{2}\left[ q_1 \pm \sqrt{q_1^2-4\zG(q_1-p_1\up{1})}\right],\\
\end{equation}
with $p_1\up{1}=\zG+(1-\zG)\upe^{-\Gz}$ and $q_1 \equiv \zG  + (1-\zG) \upe^{-\Gz} \cosh(2\Go)$.
It is easy to check that $p_1^{(1)}(\sG)=p_1^{(0)}(\sG)=p_1(\sG)$, and that $p_m^{(1)}(\sG)=p_m^{(0)}(\sG)$ when $\Go$ is set to zero. 

For $\sG=\sP$, $\lambda_-=0$, and Eq.~\eqref{eq:pm1Class} simplifies to
\begin{equation}
p_m^{(1)}(\sP)=\frac{1}{\cosh\bigl(2\Go\bigr)}{\left[\upe^{-\Gz}\cosh\bigl(2\Go\bigr)\right]}^m.
\end{equation}
Comparing $p_m^{(1)}(\sP)$ with $p_m^{(0)}(\sP)=\bigl[\upe^{-\Gz}\bigr]^m$, we see that the decay constant is modified by the additional hyperbolic cosine factor ($\geq1$).
For $\sG=\sR$ and $\sC$, both $\lambda_+$ and $\lambda_-$ are nonzero (provided $\Go\neq 0$), and Eq.~\eqref{eq:pm1Class} suggests a two-decay-rate model, a deviation from the single exponential decay behavior  observed in the time-local case. 

The two-decay-rate behavior for $\sR$ and $\sC$, however, may be difficult to observe experimentally. For $|\Go|\ll1$, we expand $\cosh(2\Go)$ in the decay rates in powers of $\Go$ to obtain
\begin{align}
\lambda_+ & = p_1\up{1} +\frac{2(p_1\up{1}-\zG)^2}{p_1\up{1}} \bigl[\Go\bigr]^2 
+O{\left(\bigl[\Go\bigr]^4\right)},\nonumber\\
\lambda_- & = \phantom{p_1\up{1}+}
\frac{2\zG(p_1\up{1}-\zG)}{p_1\up{1}} \bigl[\Go\bigr]^2
+O{\left(\bigl[\Gamma^{(1)}\bigr]^4\right)}.\label{eq:shiftClass}
\end{align}
For weak noise, $|\Gz|$ is expected to be small so that $p_1\up{1}$ is close (though smaller than) to 1, and hence $\lambda_+\sim 1$. On the other hand, $|\Go|\ll 1$, so $\lambda_-$ is much smaller in magnitude than $\lambda_+$. Thus, we have $0\lesssim \lambda_-\ll \lambda_+\lesssim 1$. Comparing with the time-local case, we see that time correlations contribute a small positive shift (relative to $p_1\up{1}$) for the $\lambda_+$ decay term, and creates a second, tiny decay term with rate $\lambda_-$. This already small $\lambda_-^m$ term quickly vanishes as $m$ grows, leaving only the dominant, more slowly decaying $\lambda_+^m$ term. One thus expects RB experiments, where $m$ is typically large, to see an exponential decay with a single decay rate, even in the presence of time correlations. This is consistent with the behavior previously observed in the numerical investigations of Ref.~\cite{Ball2016} and \cite{Epstein2014}.

One should keep in mind that the zeroth- and first-order expressions above are accurate only up to the truncation done earlier. High powers of $\Gz$ and $\Go$ kept in our expressions above may be less important than the $\Gamma\up{n}$ terms already dropped earlier, and should be discarded. This, however, cannot be done in general without fully specifying the relative magnitudes of the various $\Gamma\up{n}$ terms, and should be performed when applied to a particular physical situation.

\section{Quantum dephasing noise}\label{sec:quant}
In the previous section, we dealt with the case where the noise on the system arises through interaction with a fluctuating field of a classical nature. Here, we examine the analogous situation of the system interacting with a quantum field. The quantum nature of the noise field endows the noise seen by the system with a richer structure that we try to understand here. 

We first examine the effect of twirling in the quantum model. The time evolution operators now span both the system and bath Hilbert space. The measurement process is represented by tracing out the bath-space operators. Moreover, since only the system space can be controlled through external gates, the twirling operation is restricted to the system-space operator. The average sequence map can be written as 
\begin{equation}\label{eq:seqQuant}
    \langle \cS_{m} \rangle_\mathrm{seq} \!=\! \Bigl\langle\!
    \tw{\sG\otimes\id}\!\bigl(\cU(t_{m},t_{m-1}) \!\bigr) \cdots
    \tw{\sG\otimes\id}\!\bigl(\cU(t_1,t_0)\!\bigr) \!\Bigr\rangle_{\!\mb},
\end{equation}
where $\avg{\cdot}_{\mb} \equiv \trB \{(\cdot) \rho_{\mathrm B}\}$, a map induced by the bath state operator $\rho_{\mathrm B}$; the ``partial twirling'' map $\tw{\sG\otimes\id}$ is nothing but the normal twirling defined by the group $\sG\otimes\id$ with the identity on the bath space.
The effect of the partial twirling $\tw{\sG\otimes\id}$ on a system(qubit)-bath map $\cE$ can be thought of in a similar way as for full twirling, by writing $\cE$ as a transfer matrix as in Eq.~\eqref{eq:transMat}, but now with each matrix entry as an operator on the bath, not just a number. The partial twirling converts, as before, the transfer matrix into something diagonal, but with bath operators as the (block-)diagonal entries, rather than numbers.

In close imitation of the classical situation, we look at the case of a single-qubit system interacting with a multi-mode bosonic field---the bath---as described by the total Hamilton operator (see, for example, \cite{Legget1987}),
\begin{equation}
\mathrm{H}=\frac{1}{2}\omega_0 Z+\sum_\mu\omega_\mu a_\mu^\dagger a_\mu+ Z\otimes\sum_\mu(g_\mu a_\mu^\dagger+ g_\mu^* a_\mu).
\end{equation}
Here, $\mathrm{H}_\mathrm{S}\equiv \frac{1}{2}\omega_0 Z$ is the bare system-only Hamilton operator; $\mathrm{H}_\mathrm{B}\equiv \sum_\mu\omega_\mu a_\mu^\dagger a_\mu$ is that for the bath; the third term in $\mathrm{H}$ describes the spin-boson interaction. The sum over $\mu$ is over all the bosonic modes considered in the model. $a_\mu(a_\mu^\dagger)$ is the annihilation(creation) operator for the $\mu$th mode which has frequency $\omega_\mu$. $g_\mu$ is the complex coupling constant between the system and the $\mu$th bosonic mode. As in the classical noise model, we consider only a coupling to the $Z$ operator on the qubit system so that our model is exactly solvable. In place of a random process $b(t)$ with specified statistics as in the classical case, the time variation of the noise as seen by the system arises because the bosonic field is itself changing in time, due to its own free evolution as well as its interaction with the system. This is the natural scenario for studying time-correlated behavior as the bosonic field holds a record of what happened to the system, due to its past interaction with it.

To solve for the joint dynamical map of the qubit and bosonic field, we work in the interaction picture defined by the free-evolution Hamilton operator, $\mathrm{H}_0\equiv \mathrm{H}_\mathrm{S}+\mathrm{H}_\mathrm{B}$. The interaction picture Hamilton operator $\mathrm{H}_{\mI}(t)$ is then
\begin{equation}
\mathrm{H}_{\mI}(t)= Z \otimes \sum_\mu \bigl(
    g_\mu a_\mu^\dagger \upe^{\upi \omega_\mu t} + 
    g_\mu^* a_\mu \upe^{-\upi\omega_\mu t} \bigr).
\end{equation}
The induced time evolution operator $U(t,t')$ can be formally solved using the Magnus expansion \cite{Magnus1954}. In our case, as the Hamiltonian commutator $[\mathrm{H}_\mI(t),\mathrm{H}_\mI(t')]=-\upi 2\sum_\mu|g_\mu|^2\sin\omega_\mu(t-t')$ is proportional to the identity operator, the third- and higher-order terms in the Magnus series---all of which contain only terms with multiply nested commutators---vanish, leaving the first two Magnus terms as the only nonzero ones. This gives
\begin{equation}\label{eq:UI1}
U_\mI(t,t')=\upe^{\upi\varphi(t-t')}\exp{\left[-\upi\int_{t'}^t\upd\tau \mathrm{H}_\mI(\tau)\right]},
\end{equation}
where 
$\varphi(t-t') = \sum_\mu \frac{|g_\mu|^2}{\omega_\mu^2} \,[\omega_\mu (t-t')-\sin\omega_\mu (t-t') ]$, which
results from the double time integral of the commutator. The $\upe^{-\upi\varphi}$ adds only a global phase, and has no dynamical consequences; we drop it from our expressions below. 
Evaluating the time integral in Eq.~\eqref{eq:UI1}, we have,
\begin{equation}\label{eq:UI2}
U_\mI(t,t') = \exp{\left[-\upi \, Z\otimes B(t,t')\,\right]},
\end{equation}
where we have defined the Hermitian bath-only operator
\begin{equation}
B(t,t')\equiv -\upi \, \sum_\mu \left( \alpha_\mu(t,t') a_\mu^\dagger - \alpha^*_\mu(t,t') a_\mu \right),
\end{equation}
with $\alpha_\mu (t,t') \equiv \frac{g_\mu}{\omega_\mu}{(\upe^{\upi\omega_\mu t}-\upe^{\upi\omega_\mu t'})}$ a time-dependent complex number.
We recognize $\upe^{\upi B(t,t')}$  as the displacement operator commonly used in bosonic problems.

With this time evolution operator $U_\mI(t,t')$, we can write down, with some straightforward algebra, the transfer matrix for the time evolution map $\cU_\mI(t,t')(\cdot)\equiv U_\mI(t,t')(\cdot)U_\mI(t,t')^\dagger$, 
\begin{equation}
\cU_\mI(t,t')={\left(
\begin{array}{cccc}
\cD_+&0&~0&\cD_-\\
0&\cX+&\!-\cX_-&0\\
0&\cX_-&~\cX_+&0\\
\cD_-&0&~0&\cD_+
\end{array}\right)},
\end{equation}
where $\cD_\pm\equiv \cD_\pm(t,t')$ and $\cX_\pm\equiv \cX_\pm(t,t')$ are linear maps on operators on the bath,
\begin{align}\label{eq:bathOp}
\cD_\pm(\,\cdot\,)&\equiv\tfrac{1}{2} 
\left[ \upe^{-\upi B}(\,\cdot\,)\upe^{ \upi B}
\pm    \upe^{ \upi B}(\,\cdot\,)\upe^{-\upi B} \right], \nonumber\\
\cX_\pm(\,\cdot\,)&\equiv \tfrac{1}{2c_\pm} 
\left[ \upe^{ \upi B}(\,\cdot\,)\upe^{ \upi B}
\pm    \upe^{-\upi B}(\,\cdot\,)\upe^{-\upi B} \right],
\end{align}
with $c_+\equiv 1$ and $c_-\equiv \upi$, and the time arguments $(t,t')$ have been suppressed above, for brevity. 
The transfer matrix above is in a mixed notation, with system operators ``vectorized" using the Pauli operator basis, but the bath operators remain as operators. More specifically, the $i,j$ entry ($i$th row, $j$th column) of the above transfer matrix is the map on the bath operators $\cU_{ij}(\,\cdot\,)\equiv \frac{1}{2}\trS\{\sigma_i U_\mI(t,t')(\sigma_j\otimes\,\,\cdot\,\,)U_\mI(t,t')^\dagger\}$, for $i=0,1,2$, and $3$.

Before we add in the RB gates, it is useful to consider the effect on the system due purely to the joint evolution with the bath, to become familiar with the quantities of interest and for comparison with the case of classical noise. 
The effect on the system alone can be obtained by tracing over the bath degrees of freedom. 
Linearity permits carrying out the partial trace in the transfer-matrix form of $\cU_\mI(t,t')$ by taking the trace (over the bath) of every matrix entry. This gives the transfer matrix of the evolution map on the system, in a similar fashion as in the classical case,
\begin{equation}\label{eq:freeQuant}
\cE_{\mathrm{free}}(t,t')= {\left(\!
\begin{array}{cccc}
\avg{1}_\mb &0&~0&0\\
0& \avg{\cos 2B}_\mb & \!\!-\avg{\sin 2B}_\mb &0\\
0&\avg{\sin 2B}_\mb & ~\avg{\cos 2B}_\mb &0\\
0&0&~0&\avg{1}_\mb
\end{array}\!\right)},
\end{equation}
where we recall the notation $\avg{A}_\mb\equiv\tr\{(\cdot) A\}$ for bath operator $A$, with the trace understood to be over the bath space since only bath operators enter here. Each entry in $\cE_\mathrm{free}$ is hence a linear functional on the bath operators, not just a simple numerical value as in the classical case. To obtain Eq.~\eqref{eq:freeQuant}, we used the cyclic property under the trace: $(\tr\circ\cD_\pm)(\cdot)=\tr\{(\cdot)\, \frac{1}{2}[\upe^{\upi B}\upe^{-\upi B}\pm\upe^{-\upi B}\upe^{\upi B}]\}$, giving $\tr(\cdot)$ for the plus sign, and $0$ for the minus sign; similar steps follow for $\cX_{\pm}$.

Further evaluation of the bath functionals requires knowledge of the bath state at initial time $t'$. A common situation is where the system and bath are initially uncorrelated, with the bath in its thermal state (defined by $\mathrm{H}_\mathrm{B}$): $\rho(t')=\rho_\mathrm{s}(t')\otimes\rho_\mathrm{B,th}$, with $\rho_\mathrm{B,th}\equiv\frac{1}{ Z_\mathrm{B} }\upe^{-\beta \mathrm{H}_\mathrm{B}}$, where $Z_\mathrm{B} \equiv\tr(\upe^{-\beta \mathrm{H}_\mathrm{B}})$, and $\beta$ is the inverse temperature. In this case, we would have $\avg{\sin 2B}_\mathrm{b}=\tr(\rho_{\mathrm{B,th}}\sin 2B)= 0$ and $\avg{\cos 2B}_\mathrm{b} \equiv \upe^{-\Gamma(t,t')}$, where
\begin{align}
&\Gamma(t,t')\!\equiv\!  2(t-t')^2\!\sum_\mu\!|g_\mu|^2{\left\{\!\sinc\!{\left[\frac{\omega_\mu(t-t')}{2}\right]}\!\right\}}^{\!2}\!\!\coth\!{\left(\!\!\frac{\beta \omega_\mu}{2}\!\!\right)}\nonumber\\
&\rightarrow \!\frac{1}{2}(t-t')^2\!\!\int_{-\infty}^{\infty}\!\!\!\upd\omega J(\omega){\left\{\!\sinc\!{\left[\frac{\omega(t-t')}{2}\right]}\!\right\}}^{\!2}\!\!\coth\!{\left(\!\!\frac{\beta \omega}{2}\!\!\right)},
\end{align} 
where in the second line, we have made the continuous-frequency replacement $2 \sum_\mu |g_\mu|^2 \rightarrow \frac{1}{2}\int_{-\infty}^\infty \upd\omega J(\omega)$ to introduce the bath spectral density $J(\omega)$.
This results in $\cE_{\mathrm{free}}=\operatorname{diag}\{1,\upe^{-\Gamma},\upe^{-\Gamma},1\}$, a dephasing channel, as in the classical case, though with a modified dephasing parameter. 

Now consider $\langle \cS_{m}\rangle_\mathrm{seq}$  according to Eq.~\eqref{eq:seqQuant} by applying gates in groups $\sG=\sP,\sR$, and $\sC$, which involves in the product of partially twirled maps $\tw{\sG\otimes\id}\bigl(\cU(t_k,t_{k-1})\bigr)$. Following the earlier argument on partial twirling, this is identical to the usual full twirling, except that the matrix entries are now maps on the bath space, rather than scalar values. We thus have
\begin{equation}\label{eq:twQuant}
\begin{aligned}
\tw{\sP\otimes\id}\bigl(\cU(t_k,t_{k-1})\bigr)&=\textrm{diag}{\left\{\cD_+,\cX_+,\cX_+,\cD_+\right\}},\\
\tw{\sR\otimes\id}\bigl(\cU(t_k,t_{k-1})\bigr)&=\textrm{diag}{\left\{\cD_+,\cV,\cX_+,\cV\right\}},\\
\tw{\sC\otimes\id}\bigl(\cU(t_k,t_{k-1})\bigr)&=\textrm{diag}{\left\{\cD_+,\cW,\cW,\cW\right\}},
\end{aligned}
\end{equation}
where $\cV\equiv \frac{1}{2}(\cD_++\cX_+)$ and $\cW\equiv \frac{1}{3}(\cD_++2\cX_+)$. Again, the explicit time arguments have been suppressed. $\langle \cS_{m}\rangle_\mathrm{seq}$ is then built from the products (compositions) of these twirled maps, followed by a trace over the bath. The resulting channel is [c.f. Eq.~\eqref{eq:twClass}],
\begin{equation}\label{eq:twQuant1}
\langle \cS_{m}\rangle_\mathrm{seq}={\left\{\begin{array}{ll}
\mathrm{diag}\{d_m,f_m,f_m,d_m\},&\sG=\sP\\
\mathrm{diag}\{d_m,g_m,f_m,g_m\},&\sG=\sR\\
\mathrm{diag}\{d_m,h_m,h_m,h_m\},&\sG=\sC
\end{array}\right.}.
\end{equation}
The decay ``parameters" $d_m, f_m, g_m$ and $h_m$ are now linear functionals on the bath operators associated with the appropriate system Pauli operator basis element,
\begin{align}
d_m(\cdot)&\equiv \tr\{[\cD_+]_m(\cdot)\}, &f_m(\cdot)\equiv \tr\{[\cX_+]_m(\cdot)\},\nonumber \\
g_m(\cdot)&\equiv \tr\{[\cV]_m(\cdot)\},  &h_m(\cdot)\equiv \tr\{[\cW]_m(\cdot)\},
\end{align}
with the shorthand $[\cF]_m \equiv \cF(t_{m},t_{m-1})\circ\ldots\circ\cF(t_1,t_0)$. 

In the classical case, we were able to handle all three groups $\sP, \sR$ and $\sC$ in a unified manner, by introducing the multiset $A_\sG$. Here, we can do something similar by noting a common structure in $\cD_+,\cX_+,\cV$ and $\cW$, namely,
\begin{equation}
\frac{1}{|A_\sG|} \sum_{(a,a')\in A_\sG} \upe^{\upi a B}(\,\cdot\,) \,\upe^{-\upi a' B} =
\left\langle \upe^{\upi a B}(\,\cdot\,) \,\upe^{-\upi a' B} \right\rangle_{(a,a')},
\end{equation}
for $a, a'$ taking values $+1$ or $-1$, depending on the group $\sG$. $A_\sG$ is now composed of 2-component tuples $(a,a')$. We can write a single expression for the decay parameters,
\begin{align}\label{eq:pmQuant}
&\quad [p_m(\sG)](\cdot)\\
&\equiv {\left\langle\tr\!{\left[\upe^{\upi a_m\! B_m}\!\ldots\upe^{\upi a_1\! B_1}(\cdot)\upe^{-\upi a_1' \!B_1}\!\cdots \upe^{-\upi a_m'\!B_m}\right]}\right\rangle}_{\!(\!\vec a_m,\vec a_m'\!)}\nonumber \\
&= \left\langle \left\langle
\upe^{-\upi a_1' \!B_1}\! \cdots \upe^{-\upi a_m'\!B_m}
\upe^{\upi a_m\! B_m}\!\cdots\upe^{\upi a_1\! B_1} 
\right\rangle_{\mb} \right\rangle_{\!(\vec a_m,\vec a_m'\!)}, \nonumber
\end{align}
where $B_k\equiv  B(t_{k},t_{k-1})$, and $\avg{\,\cdot\,}_{(\!\vec a,\vec a'\!)}$ denotes the average over $(\vec a,\vec a')\in A_\sG^m$, where $\vec a_m\equiv (a_1,\ldots, a_m)$, $\vec a_m'\equiv (a'_1,\ldots, a'_m)$, and $(a_k,a'_k)\in\! A_\sG$, for $A_\sG$ for different $\sG$s as defined in Table \ref{tab:AGQuant}. With this notation, we have $p_m(\sId)$, $p_m(\sP)$, $p_m(\sR)$, and $p_m(\sC)$ for $d_m$, $f_m$, $g_m$, and $h_m$, respectively.

\begin{table}[ht]
\renewcommand*{\arraystretch}{1.5}
\centering
\begin{tabular}{c|r|c|c} 
$\sG$\  & $A_\sG$\qquad \ & \ $|A_\sG|$  &decay parameter $p_m(\sG)$  \\[0.5ex] 
\hline\hline
$\sId$ &$\{(1,1),(-1-1)\}$&2&$d_m$\\
\hline
$\sP$ & $\{(1,-1),(-1,1)\}$ & 2 &$f_m$\\ 
\hline
$\sR$ & $\{A_\sId,A_\sP\}$ & 4  &$g_m$ \\
\hline
$\sC$ & $\{A_\sId,A_\sP,A_\sP\}$ & 6 &$h_m$ \\
\end{tabular}
\caption{The multisets $A_\sG$ for the quantum noise example.}
\label{tab:AGQuant}
\end{table}

All that remains is to evaluate $p_m(\sG)$. Here, one needs to be careful that the $B_k$ operators do not commute. The commutator $[B_j,B_k]\equiv -\upi \Phi^{(j,k)}$, however, is a simple scalar: $\Phi^{(j,k)}= 2\sum_\mu\mathrm{Im}{[\alpha_\mu(t_{j},t_{j-1})\alpha^*_\mu(t_{k},t_{k-1})]}$. This simplifies the problem significantly, allowing combination of the exponential factors in $p_m(\sG)$, while picking up only an extra phase.

To proceed, we assume, as in the classical case, the RB gates are applied at constant time intervals $\Delta t=t_k-t_{k-1}~\forall k$. $\Phi^{(j,k)}$ then depends only on the difference $j-k$, and simplifies to $\Phi^{(j,k)}=\Phi^{(j-k)}=\mathrm{sgn}(k-j)\Phi^{(n)}$ for $n\equiv |j-k|$, with $\mathrm{sgn}(0)\equiv 0$. Here, 
\begin{align}\label{eq:PhiQuant}
\Phi^{(n)}&\equiv2(\Delta t)^2\sum_\mu|g_\mu|^2{\left\{\!\sinc{\left[\frac{\omega\mu\Delta t}{2}\right]}\!\right\}}^{\!2}\!\sin(n\omega_\mu\Delta t)\nonumber\\
&\rightarrow\frac{1}{2}\int_{-\infty}^\infty \upd x S(x){\left[\sinc{\left(\frac{x}{2}\right)}\right]}^2\sin(nx),
\end{align}
where we have again made the replacement $2 \sum_\mu |g_\mu|^2 \rightarrow \frac{1}{2}\int_{-\infty}^\infty \upd\omega J(\omega)$, and a further substitution for dimensionelss variables: $x\equiv \omega \Delta t$ and $S(x)\equiv (\Delta t)J(\tfrac{x}{\Delta t})$, in close imitation of what was done in the classical case.

Applying the Baker-Campbell-Hausdorff formula \cite{Hall2015} repeatedly to combine the exponential factors, we have
\begin{equation}\label{eq:Bs}
\upe^{\upi a_m B_m}\!\ldots\upe^{\upi a_2 B_2}\upe^{\upi a_1 B_1}=\upe^{\upi\vec a_m\cdot\vec B_m}\exp\biggl[ \frac{\upi}{2}\!\sum_{j,k; j>k}  \!\!\!\!a_j a_k\Phi\up{j-k}\biggr],
\end{equation}
where $\vec B_m\equiv (B_m,\ldots,B_2, B_1)$.
Applying this formula to the $a_k$ and $a_k'$ exponentials in Eq.~\eqref{eq:pmQuant} leads to a product of $\upe^{\upi\vec a_m\cdot\vec B_m}$ and $\upe^{-\upi\vec a_m'\cdot \vec B_m}$, along with two extra phase factors. Combining these two final exponential factors yields, after a little algebra, 
\begin{equation}
\upe^{\upi(\vec a_m- \vec a_m')\cdot \vec B_m}
\exp\biggl[\,\frac{\upi}{2} \!\!\sum_{j,k;j>k}\!\! (a_j-a'_j)(a_k+a'_k) \Phi\up{j-k} \biggr],
\end{equation}
as the quantity to be averaged in Eq.~\eqref{eq:pmQuant}. 

Now, since only the $(a-a')$ and $(a+a')$ combinations appear in the expression above, it is natural to change variables from $(a_k, a_k')$ to $(u_k,v_k)\equiv\frac{1}{2}(a_k-a_k',a_k+a_k')$. The original multiset $A_{\sG}$ of $(a,a')$ pairs are accordingly replaced by a new multiset $V_{\sG}$ of $(u,v)$ pairs. We also define the zero ratio $z_G$ for $V_{\sG}$ as the number of elements whose first component (the $u$) is 0, divided by the size of $V_\sG$. These various quantities are summarized in Table~\ref{tab:VGQuant}, and $p_m(\sG)$ can be written as 
\begin{equation}\label{eq:pmQuant1}
p_m(\sG) = \left\langle\!\left\langle \upe^{\upi 2 \vec u_m \cdot \vec B_m} \right\rangle_\mb
\upe^{\upi 2 \sum_{j>k} u_j v_k \Phi\up{j-k} }\right\rangle_{\!(\!\vec u_m,\vec v_m)}.
\end{equation}

\begin{table}[hbt]
\renewcommand*{\arraystretch}{1.5}
\centering
\begin{tabular}{c|r|r|c|c|c} 
$\sG$ & $A_\sG=\{(a,a')\}$ & $V_\sG=\{(u,v)\}$&$|V_\sG|$  &$\zG$& $p_m(\sG)$  \\[0.5ex] 
\hline\hline
$\sId$ &$\{(1,1),(-1-1)\}$&$\{(0,1),(0,-1)\}$&2&1&$d_m$\\
\hline
$\sP$ & $\{(1,-1),(-1,1)\}$ &$\{(1,0),(-1,0)\}$& 2 &0&$f_m$\\ 
\hline
$\sR$ & $\{A_\sId,A_\sP\}$ &$\{V_\sId,V_\sP\}$& 4  &1/2&$g_m$ \\
\hline
$\sC$ & $\{A_\sId,A_\sP,A_\sP\}$ &$\{V_\sId,V_\sP,V_\sP\}$& 6 &$1/3$&$h_m$ \\
\end{tabular}
\caption{The new multisets $V_\sG$ for the quantum noise example, replacing the earlier $A_\sG$ sets.}
\label{tab:VGQuant}
\end{table}

For $V_\sId$, $u=0$ for all elements, and Eq.~\eqref{eq:pmQuant1} simplifies to $p_m(\sId)=\avg{1}_\mb$, consistent with our expectation of TP maps. The expression for $p_m(\sP)$ also simplifies, since all $v$s in $V_\sP$ are zero,
\begin{equation}
p_m(\sP)=\left\langle\!\left\langle
\upe^{\upi 2\vec u_m \cdot \vec B_m}
\right\rangle_\mb\right\rangle_{\vec u_m},
\end{equation} 
where the $\vec u_m$ is understood to vary over the $u$ values only in $V_\sP^m$. Incidentally, this closely resembles the expression in the classical case Eq~\eqref{eq:pmClass}. Such similarity is, however, limited only to the Pauli group case; for the $\sR$ and $\sC$ groups, the phase factor $\upe^{\upi 2\sum_{j>k} u_j v_k \Phi\up{j-k} }$ has effect, the consequence of the non-commuting nature of the quantum mechanical operators.

Calculating the bath expectation in Eq.~\eqref{eq:pmQuant1} requires knowledge of the initial system-bath state. We again assume an initially (at time $t_0$) uncorrelated state with the bath in the thermal state, $\rho(t_0)=\rho_s(t_0)\otimes\rho_\mathrm{B,th}$.
With this initial state, we define a function $\Gamma^{(n)}$, reminiscent of $\Phi\up{n}$ of Eq.~\eqref{eq:PhiQuant},
\begin{equation}\label{eq:GammaQuant}
\Gamma^{(n)}\!= \!\frac{1}{2}\!\int_{-\infty}^\infty \!\!\!\upd x\, S(x) \coth\!{\left(\!\frac{\beta x}{2\Delta t}\!\right)}\!{\left[\sinc{\left(\frac{x}{2}\right)}\right]}^2\!\!\!\cos(nx),
\end{equation}
where we have done the same replacement, as for $\Phi^{(n)}$, by continuous frequency $\omega$ and then used the dimensionless variable $x\equiv\omega\Delta t$. Then, we can write
\begin{equation}\label{eq:therm-exp}
    \left\langle 
    \upe^{\upi 2 \vec u_m \cdot \vec B_m} \right\rangle_{\mathrm{b}} =
    \exp\biggl[-\!\!\sum_{j,k=1}^m \!u_j u_{k} \,\Gamma^{(|j-k|)}\biggr].
\end{equation}
In deriving this expression, we have used the standard result for the thermal expectation value of a displacement operator \cite{breuer2002theory},
\begin{equation}
\avg{\upe^{\eta_\mu a_\mu^\dagger-\eta_\mu^* a_\mu}}_\mathrm{b} = \exp\left[-\frac{1}{2} |\eta_\mu|^2 
\coth{\left(\frac{\beta \omega_\mu}{2}\right)}\right],
\end{equation}
and observed that $\upe^{\upi2\vec u_m\cdot\vec B_m}=\prod_\mu\upe^{\eta_\mu a_\mu^\dagger-\eta_\mu^* a_\mu}$ with $\eta_\mu\equiv 2\sum_k u_k\alpha_\mu(t_k,t_{k-1})$.
Substituting this back into Eq.~\eqref{eq:pmQuant1}, the decay parameter is now given by
\begin{equation}\label{eq:pmQuantFull}
p_m\!(\sG)\!= \!\left\langle\!
\upe^{-\!\sum_{jk} \!u_j u_{k} \,\Gamma^{(|j-k|)}+
\upi 2 \sum_{j>k} \! u_j v_k \Phi\up{j-k} }
\!\right\rangle_{\!\!(\!\vec u_m,\vec v_m\!)}.
\end{equation}
The seemingly complex nature of $p_m(\sG)$, because of the imaginary $\upi$ in Eq.~\eqref{eq:pmQuantFull}, is only apparent; it turns into a real cosine upon averaging over $(\vec u_m, \vec v_m)$, as we see below.

As in the classical case, we seek approximations of the exact $p_m(\sG)$ by truncating the sum in  Eq.~\eqref{eq:pmQuantFull} to include small $n\equiv |j-k|$ values only. $\Gamma^{(n)}$ and $\Phi^{(n)}$ have similar integrands, both containing $S(x)$ together with an oscillatory term, $\cos(n x)$ for $\Gamma^{(n)}$ and $\sin(n x)$ for $\Phi^{(n)}$.
We expect $\Gamma^{(n)}$ and $\Phi^{(n)}$ to be comparable in magnitude for different $n>0$, especially at low temperatures for which $\coth(\beta x/2\Delta t)\simeq 1$ in $\Gamma^{(n)}$. We assume a similar hierarchy of terms as in Eq.~\eqref{eq:hiClass}:
\begin{equation}\label{eq:hiQuant}
    1\gg |\Gamma^{(0)}| > |\Gamma^{(1)}|\sim|\Phi^{(1)}|\gg |\Gamma^{(2)}|\sim|\Phi^{(2)}|\gg\cdots.
\end{equation}
With this hierarchy, we can obtain approximations to $p_m(\sG)$, through similar steps as used in the classical case.

The zeroth-order approximation obtained by keeping only $n=0$ is no different from the classical case, 
\begin{equation}
p_m\up{0}(\sG) =\bigl[\zG + (1-\zG)\upe^{-\Gz} \bigr]^m = [p_1(\sG)]^m.
\end{equation}
Keeping up to $n=1$, we have,
\begin{align}\label{eq:quantOrder1}
&p_m(\sG)\approx p_m\up{1}(\sG) \nonumber \\ 
&= \left\langle\!
\upe^{- \Gamma\up 0 |\vec u_m|^2  -2 \Gamma\up 1 \!\sum_{k=1} \!u_k u_{k+1} \, -
\upi 2 \Phi\up{1} \sum_{k} v_k u_{k+1}  }
\right\rangle_{\!(\!\vec u,\vec v\!)} \nonumber \\
&= \frac{p_1-\lambda_-}{\lambda_+-\lambda_-} \lambda_+^m + \frac{\lambda_+-p_1}{\lambda_+-\lambda_-}\lambda_-^m,
\end{align}
where the detailed steps to derive the last line is described in Appendix \ref{app:fo-q}.
This result looks similar to the classical noise case Eq.~\eqref{eq:pm1Class}, except that
$\lambda_{\pm}$ are defined differently. The exact definitions for $\lambda_{\pm}$ are given in Eq.~\eqref{eq:lambdaQuant} in Appendix \ref{app:fo-q}. Here, we are interested in small deviations from the time-local' case, for which the decay rates can be written approximately as
\begin{align}
    \lambda_{+} &\simeq p_1 + \frac{2(p_1-\zG)^2}{p_1} 
    \Bigl[(\Gamma\up 1)^2 - \frac{\zG}{(p_1-\zG) } (\Phi\up 1)^2 \Bigr]  \nonumber\\
    \lambda_{-} &\simeq \phantom{ p_1 +} \frac{2\zG (p_1-\zG)}{p_1}
    \Bigl[(\Gamma\up 1)^2 + (\Phi\up 1)^2 \Bigr].
\end{align}
We thus see that $0\lesssim \lambda_-\ll \lambda_+\lesssim 1$ holds here, just as in the classical case. The decay curve will again predominantly look like an exponential decay at the dominant rate of $\lambda_+$.
The main difference from the classical case is that the shift in the decay rate $p_1$ due to the time correlations can be either upwards or downwards for the $\sR$ and $\sC$ groups.

\section{Validity of the truncation}\label{sec:discussion}
In the previous sections, we developed a framework to predict the decay parameter $p_m(\sG)$ for different groups, under the time-correlated dephasing noise model. In principle, knowing the bath spectral power function $C(\omega)$ or spectral density $J(\omega)$, one can calculate the corresponding decoherence functions $\Gamma\up{n}$ and $\Phi\up{n}$. We then feed this information into the expressions for $p_m(\sG)$ in Eq.~\eqref{eq:pmClassFull} and \eqref{eq:pmQuantFull}. The full expression for $p_m(\sG)$ is, however, generally not solvable exactly, and we looked at approximations by truncating the sum in the full $p_m(\sG)$. Here, we explore when the truncation of the full sum over $n\equiv |j-k|$ in $p_m(\sG)$ [Eqs.~\eqref{eq:pmClassFull} and \eqref{eq:pmQuantFull}] is a good approximation.

In general, whether the truncation gives a good approximation depends on the validity of the hierarchy of decoherence functions as expressed in Eq.~\eqref{eq:hiClass} or ~\eqref{eq:hiQuant}. 
To facilitate the discussion, we focus on the zero temperature limit and note the structural similarities in the decoherence functions,
\begin{equation}\label{eq:decfunUnified}
\left\{
\begin{aligned}
\Gamma\up{n} \\
\Phi\up{n}
\end{aligned}
\right\}= \frac{1}{2}\int_{-\infty}^\infty\!\!\! \upd x\, S(x) \,{\left[\sinc{\left(\frac{x}{2}\right)}\right]}^2 
\left\{
\begin{aligned}
\cos(n x) \\
\sin(n x)
\end{aligned}
\right\},
\end{equation}
where $\Gamma^{(n)}$ is relevant for both the classical and quantum models while $\Phi^{(n)}$ appears only in the quantum case [c.f. Eqs.~\eqref{eq:GammaClass}, \eqref{eq:GammaQuant}, and \eqref{eq:PhiQuant}]. Eq.~\eqref{eq:decfunUnified} suggests that the desired hierarchy of the decoherence functions arises if there is some degree of smoothness for the spectral function $S(x)$: For a smooth enough $S(x)$, the presence of the oscillatory sine or cosine functions in the integrand leads to a smaller and smaller integral value as $n$ increases. It is interesting to see how our conclusions might change if the hierarchy of Eq.~\eqref{eq:hiClass} or ~\eqref{eq:hiQuant} fails to hold, so that our earlier zeroth- and first-order approximations no longer work well. For concreteness, we focus on Pauli group twirling for classical noise; similar conclusions hold for the other situations discussed in this work.

We first examine a simple though somewhat unrealistic case, that of DC (quasi-static) noise~\cite{Ball2016}, before considering a more physically relevant version later. DC noise refers to the situation where $\Gamma^{(n)}=\eta$ for all $n$, where $\eta$ is a constant and $0<\eta\ll 1$. This can be realised by a spectral function of the form $S(x)=\eta\delta(x)$, and describes a (classical) noise model with infinite-length correlation time, or, equivalently, $\avg{ b(t) b(t') }_\mb$ is constant for all $|t-t'|$.

\begin{figure}[ht]
\includegraphics[width=\columnwidth]{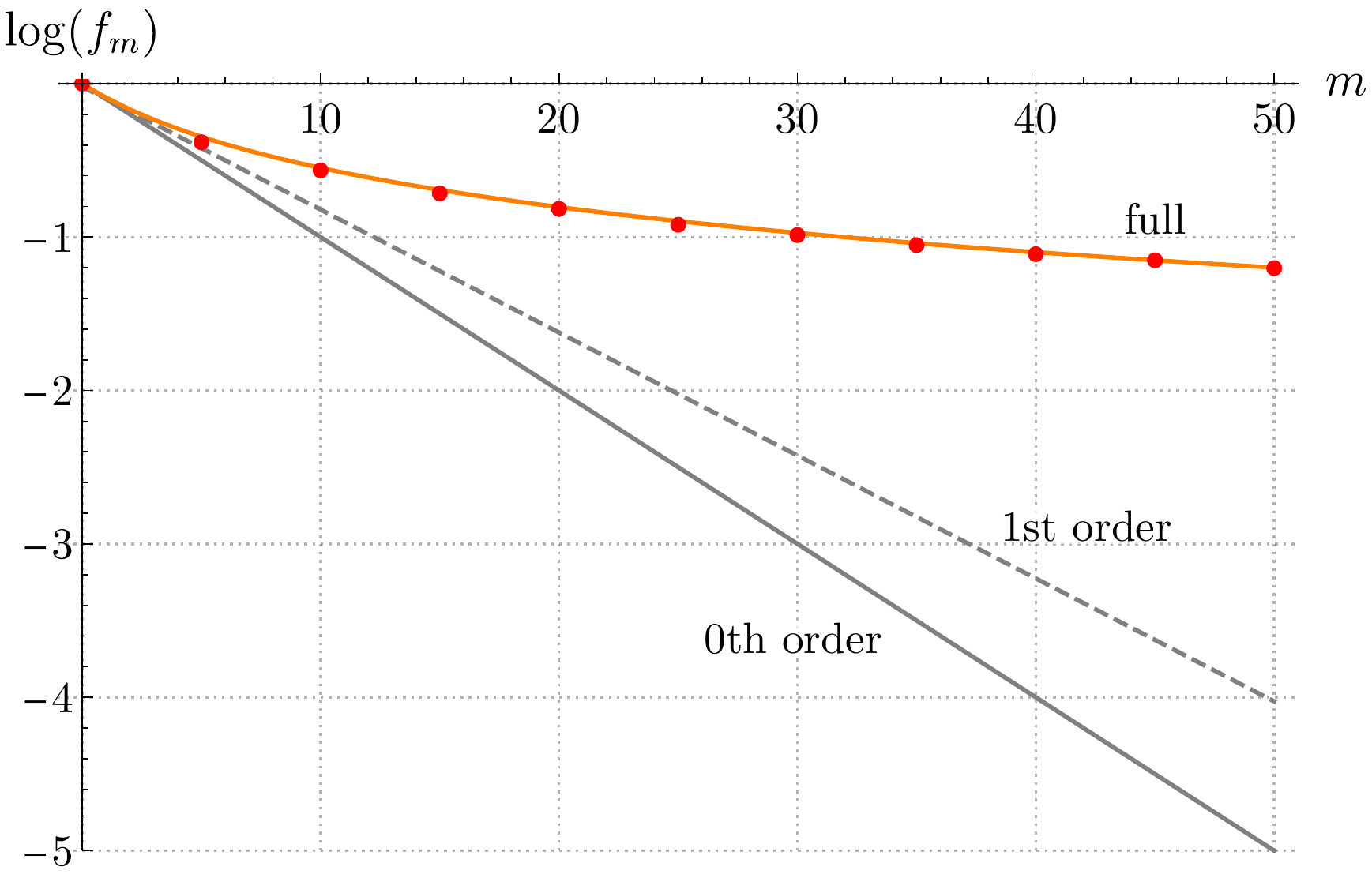}
\caption{\label{fig:quasi}The decay parameter for DC noise: $\log f_m$ versus $m$ with $\eta=0.1$. The black dashed line is the zeroth-order approximation of $f_m$; the gray solid straight line is the first-order approximation; the red dots are from Monte-Carlo averaging of Eq.~\eqref{eq:fm-qs} while the orange solid curve through the dots plots Eq.~\eqref{eq:fm-qs-app}.}
\end{figure}

In this case, the decay parameter $f_m=p_m(\sP)$ relevant for the Pauli group twirling simplifies to, without any truncation,
\begin{equation}\label{eq:fm-qs}
    f_m = \left\langle \upe^{-\eta\, \left(\sum_{k} a_k\right)^2 } \right\rangle_{\vec a_m}.
\end{equation}
Recall that $\vec a_m\in A_\sP^m$, and $A_\sP=\{-1,1\}$ (classical case, see Table \ref{tab:GClass}). if $\vec a_m$ has $m_+$ $(+1)$'s, and hence $m_-=m-m_+$ (-1)'s, then, $\sum_k a_k = 2m_+-m$. Hence, remembering the combinatorial factors, we have
\begin{equation}\label{eq:fm-qs-app}
    f_m = \frac{1}{2^m} \sum_{m_+=0}^m \binom{m}{m_+} \upe^{-\eta(2m_+-m)^2}\approx\frac{1}{\sqrt{1+2\eta m}},
\end{equation}
where the last expression is  obtained by replacing the binomial sum by an integration over the corresponding normal distribution.  These approximations are formally valid only for large enough  $m$, but in practice, work very well for $m\gtrsim 4$. As shown in Fig.~\ref{fig:quasi}, the decay curve (solid line with dots) deviates significantly from the exponential decay behavior predicted by the zeroth-order model; the first-order model does a similarly poor job of replicating the shape of the exact decay curve.

\begin{figure}[htb]
\centering
\includegraphics[width=\columnwidth]{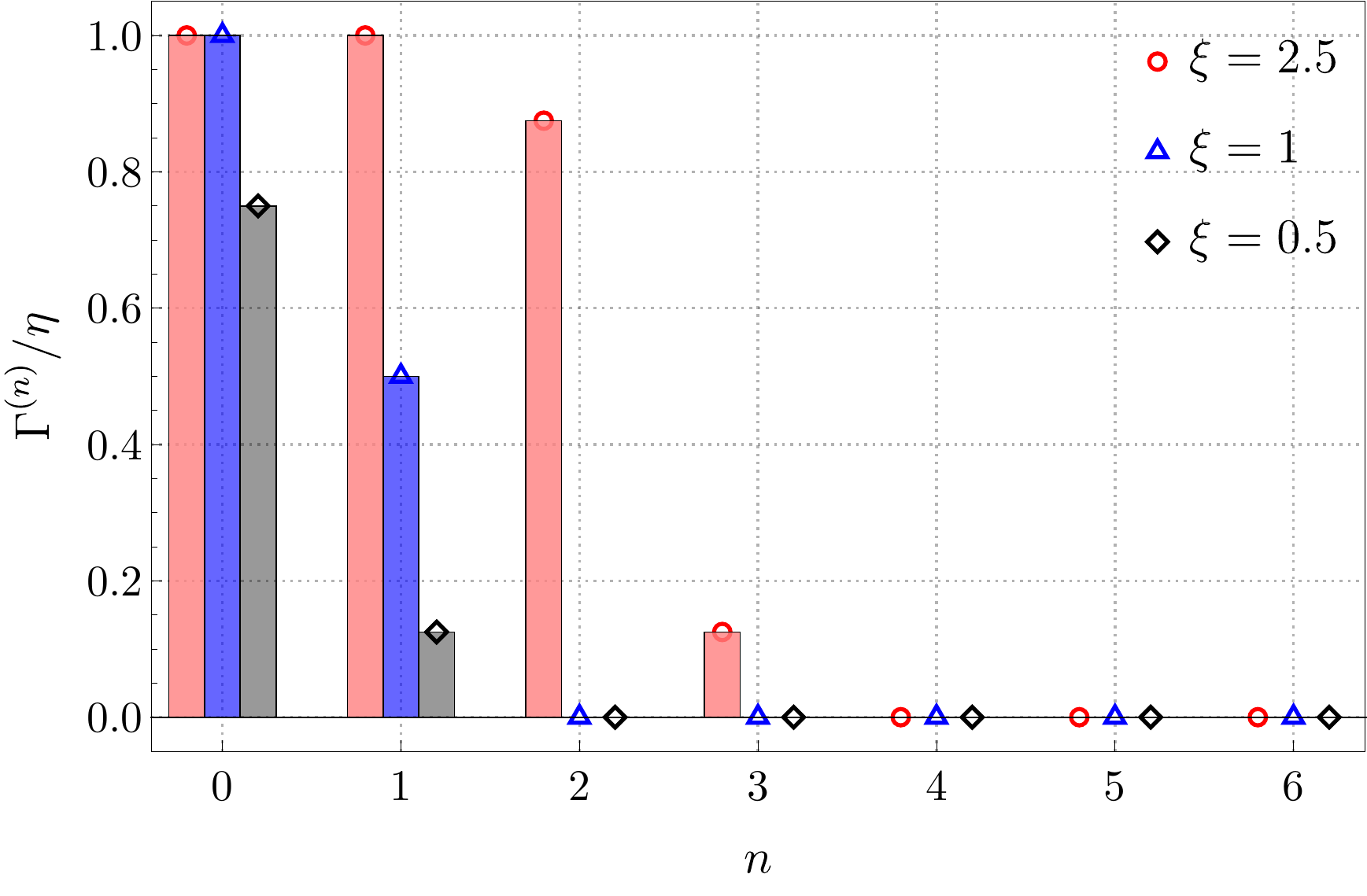}
\caption{\label{fig:corrtime}The decoherence functions $\Gamma\up{n}$ measured in units of $\eta$ for a rectangular correlation function [Eq.~\eqref{eq:RectC}] with correlation time specified by $\xi =0.5,1$, and $2.5$.  Observe that $\Gamma^{(1)}$ is non-zero in all three cases, and $\Gamma^{(n)}$ vanishes for $n\geq \xi+1$.
}
\end{figure}

The DC noise example involves infinitely long, and hence unrealistic, correlation time. We can instead consider a modified version with finite correlation time. The relevant time scale here, set by the RB pulses, is $\Dt$, a fixed quantity. Let us assume the following rectangular correlation function,
\begin{equation}\label{eq:RectC}
C(t) = {\left\{\begin{array}{ll}
2 \eta/(\Dt)^2 & \quad \textrm{for } |t/\Dt|\le \xi, \\
0 & \quad \textrm{otherwise,}
\end{array}
\right.}
\end{equation}
where $\eta>0$ and $\xi>0$ are dimensionless parameters characterizing the (dimensionless) correlation strength and correlation time, respectively. A smooth cutoff can be added to match more realistic situations, but for analytical simplicity, we adopt a sharp cutoff at $|t/\Dt|=\xi$. 
Such a $C(t)$ gives spectral function $S(x)=2 \eta \sin(\xi x)/ \pi x$. The integrals for the decoherence functions can be evaluated by complex analysis techniques, yielding the expression
\begin{align}\label{eq:corr-decfun0}
    \Gamma\up{n}&=\frac{1}{4}\eta{\left\{1+(n+\xi)[2-(n+\xi)]+[n-(1-\xi)]_{\pm}\right.}\nonumber\\
    &\quad {\left.+2[n-\xi]_\pm-[n-(\xi-1)]_\pm-[n-(\xi+1)]_\pm\right\}},
\end{align}
where we have used the shorthand $[z]_\pm\equiv z^2\,\mathrm{sgn}(z)$. The expression for $\Gamma\up{n}$ can be simplified considerably by splitting into two cases for the value of $\xi$: If $\xi\geq1$, then,
\begin{equation}\label{eq:corr-decfun}
\frac{\Gamma\up{n}}{\eta}=\!{\left\{
\begin{aligned}
    &1&& 0 \leq  n \leq \xi-1\\
    &\tfrac{1}{2}-\tfrac{1}{2}\xi_0(2-|\xi_0|)&& n=\xi+\xi_0, \textrm{ for }|\xi_0|<1\\
    &0&& n \geq \xi+1
\end{aligned}
\right.};
\end{equation}
if $0<\xi<1$, then
\begin{equation}\label{eq:corr-decfun2}
\frac{\Gamma\up{n}}{\eta}=\!{\left\{
\begin{array}{ll}
\xi(2-\xi)&\quad n=0\\
\frac{1}{2}\xi^2&\quad n=1\\
0&\quad n\geq 2
\end{array}\right.}.
\end{equation}
Fig.~\ref{fig:corrtime} illustrates the behavior of $\Gamma\up{n}$ for different values of correlation time $\xi$.

Eqs.~\eqref{eq:corr-decfun} and \eqref{eq:corr-decfun2} lead to two observations. First, if the noise correlation time is $\xi \Dt$, there is no correlation between any pair of random (noisy) gates $\cG_k$ and $\cG_{k+n}$ separated $n\ge\xi+1$ apart---the connecting decoherence function vanishes. Put differently, our earlier truncation to first-order approximation is exact if the correlation time in the system is smaller than or equal to the gate interval $\Delta t$ so that we have $\Gamma\up{n\geq 2}=0$. If not, we expect to see the non-exponential decay observed in Fig.~\ref{fig:quasi}. Second, the first-order correction $\Gamma\up{1}$ is important, at least for this example: It is never zero, or even small compared to $\Gamma\up{0}$ (see Fig.~\ref{fig:corrtime}). This $\Gamma\up{1}$ correction is altogether absent in the conventional RB treatment. 

As another example, one could also imagine a ``spiky'' spectral function where $S(x)$ is a delta function at some $x_c\neq 0$ [which does not happen to coincide with the root of $\sinc(x/2)$ or $\cos(nx)$], or is a sum of such delta spikes. Both would lead to a violation of the hierarchy Eq.~\eqref{eq:hiClass} or ~\eqref{eq:hiQuant}. Such scenarios can arise in the classical noise model if the system is driven by a highly regular perturbation; or, in the quantum model, the bath mode has only a few sharp transition frequencies.
In contrast to the spiky-spectrum example is the white-noise case, where $S(x)=\eta/\pi$, a constant. In this case, the decoherence function is nonzero only for the ``time-local'' one, i.e.,
\begin{equation}
\Gamma\up{n} ={\left\{ \begin{array}{ll}
  \eta &\quad n=0\\
  0 &\quad n\ge1  
\end{array}\right.}.
\end{equation}

A more realistic example lies in the middle ground between the spiky-spectrum and the white-noise examples, namely, $S(x)$ comprises peaks with finite spreads.
As an example, we consider an $S(x)$ built from two symmetrically placed Gaussian peaks at $\pm \bar x$ \footnote{The reason for using two Gaussian peaks instead of one comes from the parity argument: $C(\tau)$ can be defined to be even, so that its Fourier transform $S(x)$ is also even, as achieved with two symmetrically placed Gaussian peaks.},
\begin{equation}\label{eq:SxFiniteSpread}
    S(x)=\frac{\eta}{\sqrt{2\pi}\Sigma} {\left[\upe^{-\frac{1}{2\Sigma^2}(x- \bar x)^2} +\upe^{-\frac{1}{2\Sigma^2}(x+ \bar x)^2} \right]},
\end{equation}
where $\Sigma>0$ characterizes the spread of each peak. This spectrum can be derived from  noise  correlation function of the form $C(\tau) \propto \upe^{-\frac{1}{2}(\sigma \tau/\Dt)^2}\cos(\bar{x} \tau/\Dt)$. Such a scenario could arise from an oscillatory driving signal that gradually dies off after some time. At the limits of 0 and infinite $\Sigma$, we cover the cases of DC and white noise, respectively.

To study the behavior of the decoherence function $\Gamma\up{n}$, we examine two regimes based on $\bar x/\Sigma$. 
The first regime is $\bar x/\Sigma \ll 1$, where the two Gaussian peaks merge into a single plateau. 
A plot of $\Gamma\up{n}$ looks similar to Fig.~\ref{fig:corrtime}---a gradually decreasing function against $n$. This similar behavior originates from the similarity between their spectral functions. Indeed, if we perform a Taylor expansion of Eq.~\eqref{eq:SxFiniteSpread} with respect to $x=0$, keeping to the second order, we have in this regime,
\begin{equation}
    S(x)/\eta\approx \sqrt{\frac{2}{\pi }}\frac{1}{\Sigma }-\frac{1}{\sqrt{2 \pi } \Sigma ^3} x^2.
\end{equation} 
Compared with the Taylor expansion for the spectral function in the previous model,
$S(x)/\eta=2 \sin(\xi x)/ \pi x\approx 2\xi /\pi - {\xi ^3 x^2}/{3 \pi } $, we see that it is reasonable to interpret $\Sigma$ as the inverse decoherence timescale $1/\xi$, where $\Sigma\gg 1$ corresponds to the white noise limit and $\Sigma\ll1$ leads to DC noise. 

\begin{figure}[ht]
\includegraphics[width=\columnwidth]{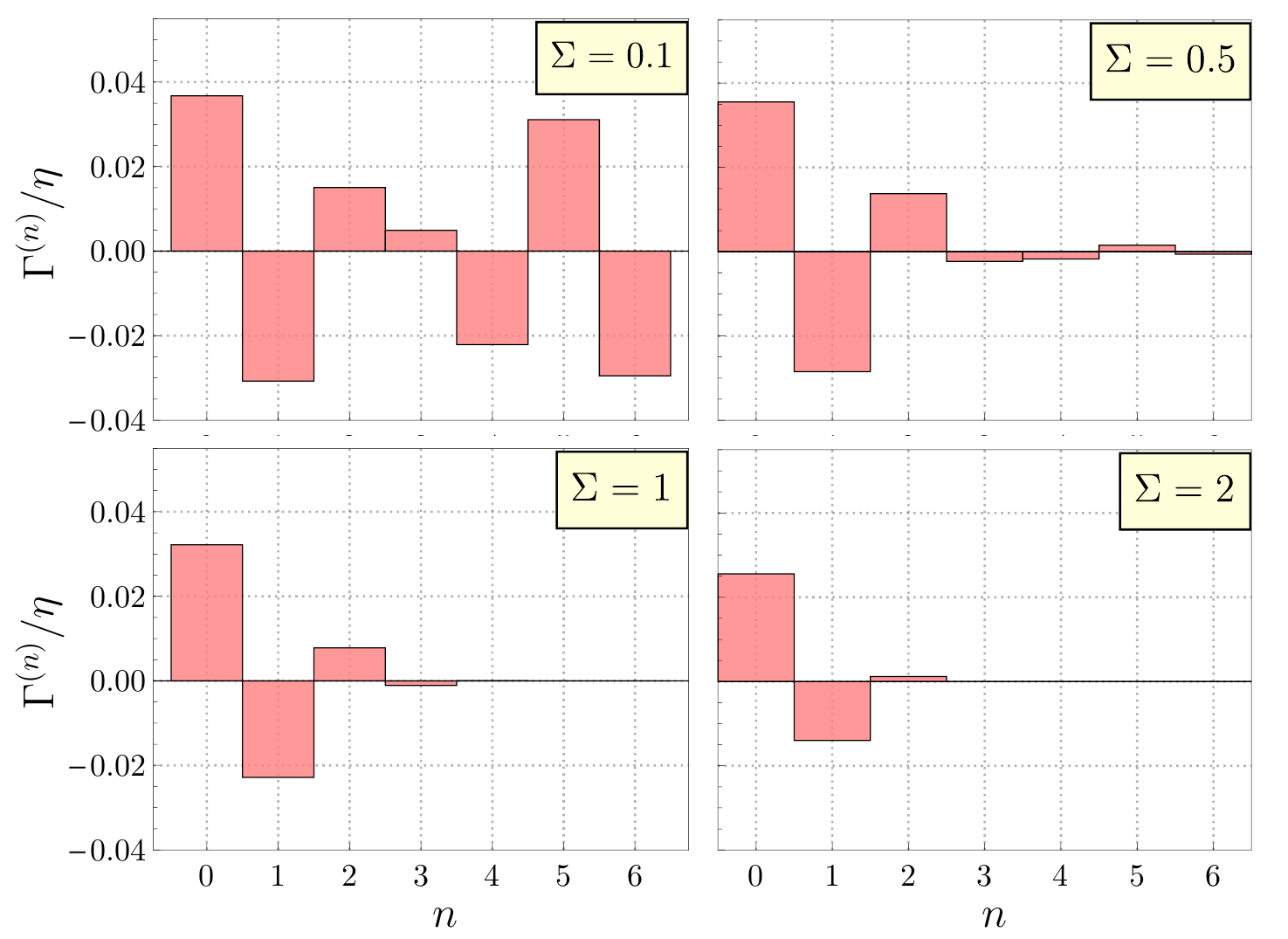}
\caption{\label{fig:finite-spread}The decoherence functions $\Gamma^{(n)}$ for a spectrum described by Eq.~\eqref{eq:SxFiniteSpread}, with $\bar x=10$ and $\Sigma$ taking on different values.}
\end{figure}

The other regime is where $\bar{x}/\Sigma> 1$; 
in this case, the behavior of $\Gamma\up{n}$ is more irregular in that it is no longer monotonically decreasing in $n$.
Numerical tests nevertheless suggest that as long as the width $\Sigma>\pi/2$, we would have $\Gamma\up{n\ge2}\approx0$. This behavior is expected as it corresponds to the intuitive picture where the spread of the peak covers at least an entire period of $\cos(nx)$ for $n\ge2$, resulting in symmetric cancellation after integrating with the oscillatory factor.  The example of $\bar x=10$, with different $\Sigma$  values, is shown in Fig.~\ref{fig:finite-spread} to illustrate this point. 

From the behavior observed in the two regimes, we conclude that the truncation at $n=1$ is a good estimate for a bath with a broad spectrum, rather than one comprising only a few narrow frequency bands. Here, the notions of ``broad" or ``narrow" are relative to the characteristic frequency scale $1/\Delta t$, set by the speed at which the gates are performed, recalling that it is $S(x)$---not $J(\omega)$ or $\widetilde C(\omega)$---that enters our analysis, where $x=\omega/\Dt$. Thus, even if $\widetilde{C}(\omega)$ or $J(\omega)$ itself is spiky, the truncation can still be a good estimation, and the exponential decay behavior maintained, by doing the gates more quickly.

\section{Conclusion}\label{sec:conc}
In this work, we examined the behavior of the randomized benchmarking protocol on qubits subjected to time-correlated dephasing noise in the classical and quantum regimes. In particular, we were interested in elucidating the circumstances for which the fidelity decay no longer follow a strict exponential law as observed in standard randomized benchmarking for static noise. The diagonal form of the Hamilton operator for the dephasing noise models permitted exact and elegant solutions to the problem, giving us the ability to examine this question without complications from approximations used in the solution. We did have to make approximations when evaluating the finals sums of Eqs.~\eqref{eq:pmClassFull} and \eqref{eq:pmQuantFull}, but these were done in a manner that respected the degree of time-correlations in the noise, directly addressing our question of the effects of those correlations. 

Our analysis here yielded expressions for the fidelity decay, retrieving the expected exponential decay as the dominant term, and offering expressions for deviations from it as an expansion in the degree of time correlations in the noise. An important general conclusion that one can draw from our expressions is that, while deviations from exponential behavior are generic---one does not have a single exponential decay as long as the $\lambda_+\neq\lambda_-$---a broad and smooth noise spectrum suffices to strongly suppress one of the decay rates compared to the other, giving the appearance of a single-rate exponential decay.

Our expressions for $p_m(\sG)$, together with the derivation techniques, while valid only for the dephasing noise situation discussed here, set the stage for further explorations of the effects of time correlations on group symmetrization methods. One could even envision making use of randomized benchmarking to characterize properties of the noise correlations, starting from detailed expressions of how those correlations modify the standard behavior.

\acknowledgments
This work is supported by a Ministry of Education Singapore Tier-2 grant (MOE2018-T2-2-142). HKN also acknowledges support by a Centre for Quantum Technologies (CQT) Fellowship. CQT is a Research Centre of Excellence funded by the Ministry of Education and the National Research Foundation of Singapore.

\appendix


\section{Explicit twirling}\label{app:twirling}
Here, we derive the explicit formulas Eqs.~\eqref{eq:twirlP}--\eqref{eq:twirlC}, for the twirling of qubit maps. We start with the map $\cE$ with transfer matrix 
\begin{equation}
\cE = {\left(\begin{array}{cccc}
    a& c_1& c_2&c_3\\ 
    b_1&x_{11} & x_{12}&x_{13}\\
    b_2 &x_{21}&x_{22}&x_{23}\\
    b_3 &x_{31}&x_{32}&x_{33}
    \end{array}\right)}.
\end{equation}
Straightforward computation gives $\tw{\sP}(\cE)=\textrm{diag}\{a,x_{11},x_{22},x_{33}\}$.

To understand the action of $\sR$-twirling, we observe that $\tw{\sR}(\cdot)=\tw{\sP}{\left(\frac{1}{2}{\left[(\cdot)+\cH(\cdot)\cH\right]}\right)}$, where $\cH$ is the Hadamard map $H(\cdot)H$. Direct computation gives
\begin{equation}
\cH\cE\cH = {\left(\begin{array}{cccc}
    a& c_3& -c_2&c_1\\ 
    b_3&x_{33} & -x_{32}&x_{31}\\
    -b_2 &-x_{23}&x_{22}&-x_{21}\\
    b_1 &x_{13}&-x_{12}&x_{11}
    \end{array}\right)},
\end{equation}
so that we have $\tw{\sR}(\cE) =\textrm{diag}\{a,v,x_{22},v\}$, with $v\equiv \tfrac{1}{2}(x_{11}+x_{33})$.

For $\sC$-twirling, we similarly observe that $\tw{\sC}(\cdot)=\tw{\sR}{\left(\frac{1}{3}{\left[(\cdot)+\cS^\dagger(\cdot)\cS+(\cS^2)^\dagger(\cdot)(\cS^2)\right]}\right)}$, where $\cS(\cdot)\equiv S(\cdot)S^\dagger$. Noting that $S\sigma_iS^\dagger =\sigma_{i+1 \textrm{ (mod 3)}}$, for $i=1,2,3$, a cyclic permutation of the Pauli operators, we see that
\begin{align}
\cS^\dagger\cE\cS &= {\left(\begin{array}{cccc}
    a& c_2& c_3&c_1\\ 
    b_2&x_{22} & x_{23}&x_{21}\\
    b_3 &x_{32}&x_{33}&x_{31}\\
    b_1 &x_{12}&x_{13}&x_{11}
    \end{array}\right)},\\
    \textrm{and}\quad
(\cS^2)^\dagger\cE(\cS^2) &= \cS\cE\cS^\dagger= {\left(\begin{array}{cccc}
    a& c_3& c_1&c_2\\ 
    b_3&x_{33} & x_{31}&x_{32}\\
    b_1 &x_{13}&x_{11}&x_{12}\\
    b_2 &x_{23}&x_{21}&x_{22}
    \end{array}\right)}.\nonumber
\end{align}
Twirling according to $\sC$ thus gives $\tw{\sC}(\cE) =\textrm{diag}\{a,w,w,w\}$, with $w\equiv \tfrac{1}{3}(x_{11}+x_{22}+x_{33})$.

\section{First-order approximations}
Here, we provide the technical derivations of the expressions for the first-order approximations of $p_m(\sG)$ for the classical and quantum noise models, as given in the main text.

\subsection{Classical noise}\label{app:class}
We describe here the details for evaluating the first-order approximation $p_m^{(1)}(\sG)$ [Eq.~\eqref{eq:classOrder1}] for the classical noise model. Eq.~\eqref{eq:classOrder1} can be equivalently written as
\begin{equation}
p_m\up{1} = {\left\langle\upe^{ -|\vec a_m|^2 \Gamma^{(0)} - 2 \Gamma^{(1)}\sum_{k=1}^{m-1} a_k a_{k+1}  }\right\rangle}_{\vec a_m},
\end{equation}
where the explicit parameter for group $\sG$ has been omitted for conciseness, and $|\vec a|\equiv \sum_{k=1}^m a_k^2$. Defining an auxiliary function $q_m(\sG)$ by
\begin{equation}
q_m = {\left\langle\cosh(2\Gamma^{(1)} a_m  )~\upe^{ -|\vec a_m|^2 \Gamma^{(0)} - 2\Gamma^{(1)} \sum_{k=1}^{m-1} a_{k} a_{k+1}}\!\right\rangle}_{\vec a_m},
\end{equation}
straightforward calculation gives
\begin{equation}\label{eq:pRec}
p_{m+1}\up{1} = \zG\, p_m\up{1} + (1-\zG) \upe^{-\Gz} q_m,
\end{equation}
and its $q_{m+1}$ counterpart,
\begin{equation}\label{eq:qRec}
q_{m+1} = z_\sG\, p_m\up{1} + (1-\zG) \upe^{-\Gz}\!\! \cosh(2\Gamma^{(1)})\, q_m .
\end{equation}
We solve the recurrence relations Eqs.~\eqref{eq:pRec} and \eqref{eq:qRec} to obtain $p_m\up{1}$. Combining the two equations into a matrix relation, we have
\begin{equation}
{\left(\!\begin{array}{c}p_{m}\up{1}\\q_{m}\end{array}\!\right)}
=M^m{\left(\!\begin{array}{c}1\\1\end{array}\!\right)},
\end{equation}
with
\begin{align}
M&\equiv {\left(\!\begin{array}{cc}
\zG & p_1\up{1}-\zG \\
\zG & q_1-\zG 
\end{array}\!\right)},\\
p_1\up{1}  &= \zG + (1-\zG) \upe^{-\Gz}, \nonumber\\
\textrm{and} \quad   q_1 &= \zG + (1-\zG) \upe^{-\Gz} \! \cosh(2\Gamma\up 1).\nonumber
\end{align}

If $\zG=0$, corresponding to the $\sG=\sP$ situation,
\begin{equation}
M^m={\left(\!\begin{array}{cc}
0& p_1^{(1)} q_1^{m-1}\\
0& q_1^m
\end{array}\!\right)},
\end{equation}
so that $p_m\up{1}=p_1\up{1} q_1^{m-1}$.  Hence, $p_m(\sP)$, in the first-order approximation, is
\begin{equation}\label{eq:appPauli}
p_m(\sP)\simeq p^{(1)}_m(\sP) = \upe^{-m\Gamma^{(0)}}{\left[\cosh\bigl(2\Gamma^{(1)}\bigr)\right]}^{m-1}.
\end{equation}
If $\zG\neq 0$, the situation for $\sG=\sR$ and $\sC$, $M$ is diagonalizable as $M=VDV^{-1}$, with $D \equiv\mathrm{diag}\{\lambda_-,\lambda_+\}$,
\begin{align}
\lambda_\pm  &= \frac{1}{2}{\left[q_1\pm\sqrt{q_1^2-4\zG \bigl(q_1-p_1\up{1}\bigr)}\right]}, \nonumber\\
\textrm{and}\quad V &= {\left(\!\begin{array}{cc}
1-\frac{\lambda_+}{\zG} &\quad 1- \frac{\lambda_-}{\zG}\\
1&1
\end{array}\!\right)}.\label{eq:appVClass}
\end{align}
Then, $M^{m}=VD^{m}V^{-1}$, and we have, after some algebra,
\begin{align}
p_m\up{1} &= \frac{p_1\up{1}-\lambda_-}{\lambda_+-\lambda_-} \lambda_+^m + \frac{\lambda_+-p_1\up{1}}{\lambda_+-\lambda_-}\lambda_-^m. \label{eq:app_qm}
\end{align}
Note that for $\sG=\sP$, we can also write our answer for $p_m^{(1)}$ in the form of Eq.~\eqref{eq:app_qm} by setting $\lambda_+=q_1$ and $\lambda_-=0$.

\subsection{Quantum noise}\label{app:fo-q}
The calculation for quantum noise proceeds in nearly exactly the same manner as that for the classical noise, except that we need to re-define the various quantities in Sec.~\ref{app:class} for the quantum situation. We have
\begin{align}
p_m\up{1}(\sG) &= {\left\langle\!\upe^{- |\vec u_m|^2\Gamma\up{0}-2\sum_{k=1}^{m-1}{\left(u_k\Gamma\up{1}-\upi v_k\Phi\up{1}\right)}u_{k+1}}\right\rangle_{\!(\!\vec u_m,\vec v_m\!)}}\nonumber\\
q_m(\sG)&\equiv{\left\langle\!\upe^{- |\vec u_m|^2\Gamma\up{0}-2\sum_{k=1}^{m-1}{\left(u_k\Gamma\up{1}-\upi v_k\Phi\up{1}\right)}u_{k+1}}\right.}\nonumber\\
&\qquad\qquad {\left.\times\cosh\!{\left[2{\left(u_m\Gamma\up{1}-\upi v_m\Phi\up{1}\right)}\right]}\right\rangle_{\!(\!\vec u_m,\vec v_m\!)}},
\end{align}
and the recurrence relations can be written as, as in the classical case,
\begin{equation}
{\left(\!\begin{array}{c}p_{m}\up{1}\\q_{m}\end{array}\!\right)}
=M^m{\left(\!\begin{array}{c}1\\1\end{array}\!\right)},
\end{equation}
but now with $M$ defined as
\begin{align}
M&\equiv {\left(\!\begin{array}{cc}
\zG & p_1\up{1}-\zG \\
\zG' & q_1-\zG' 
\end{array}\!\right)},\\
z_\sG'&\equiv \zG\cos(2\Phi\up{1}),\nonumber\\
p_1\up{1}  &= \zG + (1-\zG) \upe^{-\Gz}, \nonumber\\
\textrm{and} \quad   q_1 &= \zG' + (1-\zG) \upe^{-\Gz} \! \cosh(2\Gamma\up 1).\nonumber
\end{align}

In the quantum case, we have $\zG=1,0,1/2$, and $1/3$, corresponding to $\sG=\sId, \sP, \sR$, and $\sC$, respectively. For $\zG=1$, i.e., $\sG=\sId$, we have $p_m\up{1}(\sId)=1$, as to be expected for a product initial system-bath state. For $\zG=0$, i.e., $\sG=\sP$, we have the same answer as in the classical situation,
\begin{equation}
p^{(1)}_m(\sP) = \upe^{-m\Gamma^{(0)}}{\left[\cosh\bigl(2\Gamma^{(1)}\bigr)\right]}^{m-1}.
\end{equation}
For $\zG\neq 0$ or $1$, i.e., $\sG=\sR$ or $\sC$, we need to diagonalize $M$, as done in Sec.~\ref{app:class}: $M=VDV^{-1}$, with $D\equiv\diag{\lambda_-,\lambda_+}$, 
\begin{align}
\lambda_\pm  &= \frac{1}{2}\biggl[q_1+\zG-\zG'\nonumber\\
&\qquad {\left.\pm\sqrt{(q_1+\zG-\zG')^2-4\bigl(\zG q_1-\zG'p_1\up{1}\bigr)}\right]}, \nonumber\\
\textrm{and}\quad V &= {\left(\!\begin{array}{cc}
\frac{\zG}{\zG'}-\frac{\lambda_+}{\zG'} &\quad \frac{\zG}{\zG'}- \frac{\lambda_-}{\zG'}\\
1&1
\end{array}\!\right)},\label{eq:lambdaQuant}
\end{align}
which reduces to the classical case if we set $\zG'=\zG$.
Then, $M^{m}=VD^{m}V^{-1}$, and we have, after some algebra,
\begin{align}
p_m\up{1} &= \frac{p_1\up{1}-\lambda_-}{\lambda_+-\lambda_-} \lambda_+^m + \frac{\lambda_+-p_1\up{1}}{\lambda_+-\lambda_-}\lambda_-^m,
\end{align}
identical in structure as the classical case, but with different expressions for $\lambda_\pm$.

\end{document}